\pdfoutput=1 
\documentclass[11pt,a4paper]{article}
\usepackage{jheppub}

\usepackage{rotating} 
\usepackage{preprintcover} 
\PreprintCoverPaperTitle{Measurements of the pseudorapidity dependence of the total transverse energy in proton-proton collisions at $\sqrt{s}=7~{\rm TeV}$ with ATLAS} 
\PreprintIdNumber{CERN-PH-EP-2012-196} 
\PreprintCoverAbstract{This paper describes measurements of the sum of the transverse energy
  of particles as a function of  particle pseudorapidity, $\eta$, in
  proton-proton collisions at a centre-of-mass energy,
  \mbox{$\sqrt{s} = 7~{\rm TeV}$} using the ATLAS detector at the Large
  Hadron Collider. 
 The measurements are performed in the region
 \mbox{$|\eta| < 4.8$} for two event classes: those requiring the presence of particles with a low transverse momentum and those requiring particles with a significant transverse momentum. In the second dataset measurements are made in the region transverse to the hard scatter.
The distributions are compared to the
 predictions of various Monte Carlo event generators, which generally
 tend to underestimate the amount of transverse energy at high
 $|\eta|$. }
\PreprintJournalName{Journal of High Energy Physics}

\RequirePackage{graphicx}
\usepackage{subfigure}

\newcommand{\imb}{\mathrm{\mu b}^{-1}}

\newcommand{\topo}{cluster}
\newcommand{\topos}{clusters}
\newcommand{\ecap}{endcap}

\newcommand{\pizero}{\pi^{0}}
\newcommand{\Pt}{p_{\rm T}}
\newcommand{\Et}{E_{\rm T}}

\newcommand{\Mgg}{m_{\gamma\gamma}}
\newcommand{\Pgg}{\pizero\rightarrow\gamma\gamma}

\newcommand{\nevt}{N_{\rm evt}}
\newcommand{\sumet}{\Sigma E_{\rm T}}
\newcommand{\sumettot}{E_{\rm T}^{\rm tot}}
\newcommand{\sumettotlong}{ \frac{1}{\sumettot}\times\frac{d\sumettot}{d|E_{\rm T}|}}
\newcommand{\sumetlong}{\frac{1}{\nevt}\frac{d\nevt}{d\sumet}}
\newcommand{\etflowlong}{\langle\frac{d^{2}\sumet}{{d\eta}{d\phi}}\rangle}
\newcommand{\etflow}{E_{\rm T}^{\rm density}}
\newcommand{\nch}{N_{\rm ch}}
\newcommand{\ptch}{p_{\rm T}^{\rm ch}}
\newcommand{\etach}{\eta^{\rm ch}}
\newcommand{\pnc}{p^{\rm ch(neutral)}}
\newcommand{\njet}{N_{\rm jet}}

\newcommand{\Etjetonetwo}{\Et^{\rm jet1(2)}}
\newcommand{\etajet}{\eta^{\rm jet}}
\newcommand{\hpp}{{\tt Herwig++}}
\newcommand{\ueseven}{{\tt UE7-2}}
\newcommand{\ambt}{{\tt AMBT1}}
\newcommand{\ambtnopcuts}{{\tt AMBT1 (no p cuts)}}
\newcommand{\auetbtwo}{{\tt AUET2B:CTEQ6L1}} 
\newcommand{\perugia}{{\tt Perugia0}}
\newcommand{\DW}{{\tt DW}}
\newcommand{\pythia}{{\tt PYTHIA}}
\newcommand{\pythiasix}{{\tt PYTHIA\,6}}
\newcommand{\pythiaeight}{{\tt PYTHIA\,8}}
\newcommand{\epos}{{\tt EPOS}}
\newcommand{\eposlhc}{{\tt EPOS LHC}}
\newcommand{\fourc}{{\tt 4C}}
\newcommand{\atwo}{{\tt A2}}
\newcommand{\atwom}{{\tt A2:MSTW2008LO}}
\newcommand{\atwoc}{{\tt A2:CTEQ6L1}}

\newcommand{\geant}{{\tt GEANT4}}
\newcommand{\cteqfive}{{\tt CTEQ\,5L}}
\newcommand{\cteqsix}{{\tt CTEQ\,6L1}}
\newcommand{\mstwlo}{{\tt MSTW2008\,LO}}
\newcommand{\mrststar}{{\tt MRST\,LO*}}
\newcommand{\mrststarstar}{{\tt MRST\,LO**}}
\newcommand{\antikt}{anti-$k_{t}$}
\newcommand{\dphijj}{\Delta\phi_{\rm jj}}
\newcommand{\tablecaption}{Measured $\sumetlong$ and systematic uncertainty breakdown for the}
\newcommand{\vardefmb}{$\nch$ refers to the number of charged particles in the event, and $\ptch$ and $\etach$ are, respectively, the $\Pt$ and $\eta$ of those particles.}
\newcommand{\vardefdj}{$\njet$ refers to the number of jets, $\Etjetonetwo$ is the $\Et$ of the (sub-)leading jet, $\etajet$ is the jet pseudorapidity, and $\dphijj$ is the azimuthal angle difference between the two leading jets.}
\newcommand{\vardef}{$\pnc$ refers to the momentum of the charged(neutral) particles used in the $\sumet$ calculation.}

\begin{document}
\title{Measurements of the pseudorapidity dependence of the total transverse energy in proton-proton collisions at $\sqrt{s}=7~{\rm TeV}$ with ATLAS}

\author{The ATLAS collaboration}

\begin{abstract}
{This paper describes measurements of the sum of the transverse energy
  of particles as a function of  particle pseudorapidity, $\eta$, in
  proton-proton collisions at a centre-of-mass energy,
  \mbox{$\sqrt{s}$ = 7 TeV} using the ATLAS detector at the Large
  Hadron Collider. 
 The measurements are performed in the region
 \mbox{$|\eta| < 4.8$}
 for two event classes: those requiring the presence of particles with a low transverse momentum and those requiring particles with a significant transverse momentum. In the second dataset measurements are made in the region transverse to the hard scatter.
The distributions are compared to the
 predictions of various Monte Carlo event generators, which generally
 tend to underestimate the amount of transverse energy at high
 $|\eta|$. 
}

\end{abstract}
\maketitle
\section{Introduction}
The main aim of the Large Hadron Collider (LHC) general-purpose detectors is to explore physics in collisions around and above the electroweak symmetry-breaking scale.
Such processes typically involve high momentum transfer, which distinguishes
 them from the dominant processes, namely low momentum transfer strong force interactions described by non-perturbative Quantum Chromodynamics (QCD).
In order to collect enough data to be sensitive to rare processes it is necessary to run the LHC at  high instantaneous luminosities, meaning that multiple proton-proton interactions are very likely to occur each time the proton bunches collide. 
 It is essential that the Monte Carlo event generators used to simulate these processes have an accurate description of the soft particle kinematics in inclusive proton-proton interactions over the entire acceptance of the LHC experiments,  such that reliable comparisons can be made between theoretical predictions and the data for any process of interest.

Protons are composite objects made up of partons, the longitudinal momentum distributions of which are described by parton distribution functions (PDFs).   When protons interact at the LHC the dominant parton-parton interaction is $t$-channel gluon exchange.
Due to the composite nature of the protons it is possible that multiple parton-parton interactions (MPI) occur in the same proton-proton interaction.
Therefore, if a hard parton-parton interaction occurs it will likely be accompanied by additional QCD interactions, again predominately low momentum $t$-channel gluon exchange.
Any part of the interaction not attributed to the hard parton-parton scatter is collectively termed the underlying event, which includes MPI as well as soft particle production from the beam-beam remnants. 
Monte Carlo event generators that simulate any hard process at the LHC must also include an accurate description of the underlying event.

At low momentum transfer, perturbative calculations in QCD are not meaningful and cross-sections cannot currently be computed from first principles. Phenomenological models are therefore used to describe the kinematics of particle production in inclusive proton-proton interactions and in the underlying event in events with a hard scatter; these must be constrained by, and tuned to, data.

This paper presents a measurement of the sum of the transverse energy, $\sumet$, of particles produced in proton-proton collisions at the LHC, using the ATLAS detector~\cite{cite:atlas}. The $\sumet$ distribution is measured 
in bins of pseudorapidity\footnote{ATLAS uses a
right-handed coordinate system with its origin at the nominal
interaction point (IP) in the centre of the detector and the $z$-axis
along the beam pipe. The $x$-axis points from the IP to the centre of
the LHC ring, and the $y$-axis points upward. Cylindrical coordinates
$(r,\phi)$ are used in the transverse ($x-y$) plane, $\phi$ being the
azimuthal angle around the beam pipe. The
pseudorapidity is defined in terms of the polar angle $\theta$ with
respect to the beamline as \mbox{$\eta=-\ln\tan(\theta/2)$.}  
}, $\eta$,  
in the range \mbox{$|\eta|<4.8$}. 
Distributions of the $\sumet$ and the mean $\sumet$ as a function of $|\eta|$ are presented.
These measurements are performed with two distinct datasets. 
The first is as inclusive as possible, with minimal event selection applied, sufficient to ensure that an inelastic collision has occurred. This is termed the {\it minimum bias} dataset and is studied in order to probe the particle kinematics in inclusive proton-proton interactions. Understanding these processes is vital to ensure a good description of multiple proton-proton interactions in runs with high instantaneous luminosity.
The second dataset requires the presence of two jets with high transverse energy, $\Et>20$~GeV,
which ensures a hard parton-parton scatter has occurred and therefore allows the  particle kinematics in the underlying event to be probed. This sample is termed the {\it dijet} dataset.
Both datasets were collected during the first LHC runs at \mbox{$\sqrt{s} = 7$~TeV} in 2010. The data samples correspond to
integrated luminosities of 7.1~$\imb$
for the minimum bias measurement\footnote{The run
dependence of the analysis was checked in a larger sample and found to be negligible.}
and 590~$\imb$ for the dijet measurement.
Such small data samples are used because the early LHC runs had a very low instantaneous luminosity ensuring 
a negligible contribution from multiple proton-proton interactions.
The larger sample for the dijet analysis is used as the cross-section for a hard scatter is significantly lower than for inclusive proton-proton interactions.

Many previous measurements of the kinematic properties of particles produced in minimum bias events~\cite{cite:mb-cdf, cite:mb, cite:mb-alice, cite:mb-cms} and in the underlying event~\cite{Acosta:2004wqa, cite:ueZ-cdf, cite:ue-tracks, ALICE:2011ac, cite:ue-cms, cite:ue-topo} were restricted to the central region of the detectors. This is because they used tracking detectors, with limited coverage, to study charged particles, or because they used only the central region of the calorimeters, where the tracking detectors could be used for calibration.
Measurements of the mean of the sum of the energy of particles as a function of $|\eta|$ in minimum bias events and in the underlying event were performed with the CMS 
forward calorimeter~\cite{cite:eflow-cms}; these were limited to the very forward region \mbox{($3.15<|\eta|<4.9$)}.
LHCb has performed measurements of charged particle multiplicities in the regions \mbox{$-2.5<\eta<-2.0$} and  \mbox{$2.0<\eta<4.5$~\cite{Aaij:2011yj}.}

The measurements described in this paper utilize the entire acceptance
of the ATLAS calorimeters, \mbox{$|\eta|<4.9$},  allowing the $\sumet$
to be probed and unfolded in the region \mbox{$|\eta|<4.8$}.
Unless otherwise stated, the central region will refer to the range \mbox{$|\eta|<2.4$} and the forward region will refer to the range \mbox{$2.4<|\eta|<4.8$}.
The measurement is performed with the ATLAS calorimeters and is corrected for detector effects so that the variables are defined at the particle-level (see Section~\ref{sec:definition}), which includes all stable particles (those with a proper lifetime greater than $3 \times 10^{-11}$~seconds).
Both the mean and distributions of the $\sumet$ are measured. This provides additional information, giving a complete picture of both inclusive proton-proton interactions and the underlying event in dijet processes, within the entire acceptance of the general purpose LHC detectors. 
The relative levels of particle production in the forward and central regions may be affected by the contribution from beam-beam remnant interactions, details of the hadronization as modelled with colour reconnection between quarks and gluons, the relative contribution from diffractive processes and the parton distribution functions in this kinematic domain.

This paper is organized as follows. 
Section~\ref{sec:definition} defines the particle-level variables.
Section~\ref{sec:mc} describes the Monte Carlo models that are used to correct the data for detector effects and to compare to the final unfolded results.
The ATLAS detector is discussed in section~\ref{sec:atlas}, the event reconstruction in section~\ref{sec:reco} and the event selection in
section~\ref{sec:selection}.
The method used to correct the data for detector effects is described in section~\ref{sec:corrections}. The systematic  uncertainties are described in section~\ref{sec:systs}. Section~\ref{sec:results} presents and discusses the final results and compares them to various Monte Carlo simulations. Finally, conclusions are given in section~\ref{sec:conclu}.

\section{Particle-level variable definitions}
\label{sec:definition}
In data, events are selected and variables defined using calibrated
detector-level quantities. Corrections for detector effects are then
applied. 
In order to compare  the corrected data with predictions from Monte Carlo event generators without passing the events through a simulation of the ATLAS detector, it is necessary to define  variables at the particle-level.
The particle-level $\sumet$ is defined at the generator level by summing the $\Et$ of all stable charged particles with momentum $p > 500$ MeV  and all stable neutral particles with  $p > 200$ MeV.
 Lower momentum particles are not included as they are unlikely to deposit significant energy in the ATLAS calorimeters.

The $\sumet$ distribution is defined as $\sumetlong$, where $\nevt$ is the number of events in the sample. It is measured in six regions: \mbox{$0.0<|\eta|<0.8$}, \mbox{$0.8<|\eta|<1.6$}, \mbox{$1.6<|\eta|<2.4$}, \mbox{$2.4<|\eta|<3.2$}, \mbox{$3.2<|\eta|<4.0$} and \mbox{$4.0<|\eta|<4.8$}. 
In addition the mean $\sumet$ over all events,  per unit $\eta$--$\phi$, is measured as a function of $|\eta|$.
This is denoted as the transverse energy density ($\etflow$) and is
defined as $\etflowlong$.
In the minimum bias measurement, the $\sumet$ includes particles at any $\phi$.
In the dijet measurement, the $\sumet$ is measured using only particles that are in the azimuthal region transverse to the hard scatter, namely 
\mbox{$\frac{\pi}{3}<|\Delta\phi|< \frac{2\pi}{3}$}, where
$\Delta\phi$ is the azimuthal separation between the leading jet and a
given particle. This region of phase space contains limited particle
production from the hard parton-parton interaction and is therefore
most sensitive to the underlying event. 

\subsection{Particle-level minimum bias event selection}
\label{sec:truth-mb-selection}
The events in the minimum bias analysis contain at least two charged particles with \mbox{$\Pt > 250$~MeV} and \mbox{$|\eta|<2.5$}, reflecting as closely as possible the requirement of a reconstructed vertex, as will be discussed in section~\ref{sec:selection}.
\subsection{Particle-level dijet event selection}
\label{sec:truth-dijet-selection}
The events in the dijet analysis contain at least two particle-level jets\footnote{A particle-level jet is built from all stable particles, excluding neutrinos and muons.}. Both the leading and sub-leading jets must have  \mbox{$\Et^{\rm jet} > $~20 GeV} and \mbox{$|\eta^{\rm jet}| < 2.5$}, reconstructed with the \antikt~\cite{cite:antikt} algorithm with radius parameter $R=0.4$.
This selection ensures that a hard scattering has occurred. A relatively small radius parameter reduces the probability of the jet algorithm collecting particles that are not associated with the hard scatter.
In order to select a well balanced back-to-back dijet system, 
the jets satisfy \mbox{$|\dphijj|> 2.5$} radians, where $\dphijj$ is the difference in azimuthal angle of the leading and sub-leading jet, and \mbox{$\frac{\Et^{\rm jet2}}{\Et^{\rm jet1}} > 0.5$}, where $\Et^{\rm jet1(2)}$ is the $\Et$ of the (sub-)leading jet. The latter requirement retains most of the dataset, but avoids topologies in which there is a large transverse energy difference between the leading and sub-leading jets.
A well balanced dijet system suppresses contributions from multijet events, allowing a clearer distinction between regions with particle production dominated by the hard scatter and by the underlying event. 
\section{Monte Carlo event generators}
\label{sec:mc}
This section describes the Monte Carlo event generator (MC) models used to correct the data for detector effects, to assign systematic uncertainties to the corrections due to the physics model, and for comparisons with the final unfolded data. The  \pythiasix~\cite{cite:pythia6}, \pythiaeight~\cite{cite:pythia8}, \hpp~\cite{cite:hpp} and \epos~\cite{Werner:2010ny} 
generators are used, with various tunes that are described below. First a brief introduction to the relevant parts of the event generators is given.

\pythiasix\ and \pythiaeight\ are general purpose generators that use the Lund string hadronization model~\cite{cite:Lund}.
In \pythiasix\ there is an option to use a virtuality-ordered or $\Pt$-ordered  parton shower, with the latter used in most recent tunes. In \pythiaeight, the
 $\Pt$-ordered parton shower is used.
The inclusive hadron-hadron interactions are described by a model that
 splits the total inelastic cross-section into non-diffractive
processes, dominated by $t$-channel gluon exchange, and diffractive
processes involving a colour-singlet exchange.  
The diffractive processes are further divided into
single-diffractive dissociation, where one of the initial hadrons remains
intact and the other is diffractively excited and dissociates, and
double-diffractive dissociation where both hadrons dissociate. Such events tend to have large gaps in particle production at central rapidity.
The smaller contribution from central
diffraction, in which both hadrons remain intact and particles are produced in the central region, is neglected.   
 The $2\rightarrow2$ non-diffractive processes, including MPI, are described by
lowest-order perturbative QCD with the divergence of the cross-section
as $\Pt\rightarrow0$ regulated with a phenomenological model. 
There are many tunable parameters that control,
among other things,
the behaviour of this regularization, 
 the matter distribution of partons within the
hadrons, and colour reconnection.
When $\Pt$-ordered parton showers are used, the MPI and parton shower are interleaved in one common sequence of decreasing $\Pt$ values. For \pythiasix\ the interleaving is between the initial-state shower and MPI only, while for \pythiaeight\ it also includes final-state showers. 
Since the $\Pt$-ordered showers and interleaving with MPI are considered to be a model improvement, the most recent \pythiasix\ tunes are made with this configuration. This is also the only configuration available in \pythiaeight.
A pomeron-based approach is used to
describe diffractive events, using (by default) the Schuler and Sj\"{o}strand~\cite{cite:schuler-sjostrand}
parameterization of the pomeron flux. 
In \pythiasix\ the diffractive dissociations are treated using
the Lund string model, producing final-state particles with limited $\Pt$. 
In \pythiaeight\ the dissociations are treated like this only for events with a diffractive system with a very low mass; in higher mass systems diffractive parton distributions
from H1~\cite{Aktas:2006hy} are used to include diffractive final states which are
characteristic of hard partonic interactions. In this case, the full
machinery of MPI and parton showers is used. This approach yields a
significantly harder $\Pt$ spectrum for final-state particles.

\hpp\ is another general purpose generator, but with a different approach: it uses an angular-ordered parton shower and the cluster hadronization model~\cite{cite:cluster}. 
It has an MPI model similar to the one used by the \pythia\ generators, with tunable parameters for regularizing the behaviour at very low momentum transfer, but does not include the interleaving with the parton showers. Inclusive hadron-hadron collisions are simulated by applying the MPI model to events with no hard scattering. It is therefore possible to generate an event with zero  $2\rightarrow2$ partonic scatters, in which only beam remnants are produced, with nothing in between them. While \hpp\ has no explicit model for diffractive processes, these zero-scatter events will look similar to double-diffractive dissociation. 

\epos\ is an event generator used primarily to simulate heavy ion and cosmic shower interactions, but which can also simulate proton-proton interactions.
\epos\ provides an implementation of a parton based
Gribov-Regge~\cite{Drescher:2000ha} theory which is  an effective, QCD-inspired field theory
describing hard and soft scattering simultaneously. 
\epos\ calculations  thus do not rely on the standard PDFs as used
in  generators like \pythia\ and \hpp.
At high parton
densities a hydrodynamic evolution of the initial state is calculated for
the proton-proton scattering process as it would be for heavy ion interactions.
The results presented here use the \eposlhc\ tune, 
which contains a parameterized
approximation of the hydrodynamic evolution. The optimal parameterization
has been derived from tuning to LHC minimum bias data.

The reference MC sample used throughout this study is the \ambt~\cite{cite:ambt} tune of \pythiasix. 
In order to check the model dependence of the data corrections, additional generators and tunes are considered. These are summarized in table~\ref{tab:tunes} along with information about the PDFs used and whether minimum bias or underlying event data at $\sqrt{s}$~=~7~TeV were used in the tune.
Of the \pythiasix\ tunes listed, only \DW\ uses the old virtuality-ordered parton shower without interleaving with MPI.
Some more recent tunes are also used to compare to the unfolded data; these are summarized in table~\ref{tab:tunes-compare}. For these more recent tunes the PDF is explicitly given in the name as there are different instances of each tune that use different PDFs and hence have different parameters.
\begin{table}
\begin{center}
	\begin{tabular}{llllll} 
          \hline \hline
          Generator     & Version   &  Tune                        & PDF                                   & \multicolumn{2}{c}{7 TeV data}  \\
                        &           &                              &                                       & MB  & UE    \\   
          \hline  
          \pythiasix\   & 6.423     &  \ambt~\cite{cite:ambt}      & \mrststar~\cite{Sherstnev:2008dm}     & yes & no \\
          \pythiasix\   & 6.423     &  \DW~\cite{Albrow:2006rt}    & \cteqfive~\cite{Lai:1999wy}           & no  & no \\
          \pythiasix\   & 6.423     &  \perugia~\cite{cite:perugia}& \cteqfive                             & no  & no \\
          \pythiaeight\ & 8.145     & \fourc~\cite{cite:fourc}    & \cteqsix~\cite{Pumplin:2002vw}        & yes & no \\
          \hpp\         & 2.5.1     & \ueseven~\cite{Gieseke:2011xy} & \mrststarstar~\cite{Sherstnev:2008dm} & no  & yes \\
         \hline \hline
	\end{tabular}
	\caption{MC tunes used to unfold the data and to determine the physics model dependent systematic uncertainty. The last two columns indicate whether the data used in the tune included 7~TeV minimum bias (MB) and/or underlying event (UE) data.}
        \label{tab:tunes}
\end{center}
\end{table}

\begin{table}
\begin{center}
	\begin{tabular}{lllllll} 
          \hline \hline
          Generator     & Version                     & Tune                        & PDF                           &  \multicolumn{2}{c}{7 TeV data}  \\
                        &                             &                             &                               & MB  & UE    \\   
          \hline  
          \pythiasix\   & 6.425                       & \auetbtwo~\cite{cite:auetb} & \cteqsix                      & no  & yes \\
          \pythiaeight\ & 8.153                       & \atwoc~\cite{cite:atwo}     & \cteqsix                      & yes & no \\
          \pythiaeight\ & 8.153                       & \atwom~\cite{cite:atwo}     & \mstwlo~\cite{Martin:2009iq}  & yes & no \\
          \epos\        & 1.99$\_$v2965               & LHC                         & N/A                           & yes & no \\ 
         \hline \hline
	\end{tabular}
	\caption{Additional MC tunes used to compare to the unfolded data only. The last two columns indicate whether the data used in the tune included 7~TeV minimum bias (MB) and/or underlying event (UE) data.}
        \label{tab:tunes-compare}
\end{center}
\end{table}

\section{The ATLAS detector}
\label{sec:atlas}
The ATLAS detector is described in detail in ref.~\cite{cite:atlas}. Here only the components most relevant for this measurement are described.

Tracks and interaction vertices are reconstructed with the 
inner detector tracking system, which consists of a silicon pixel detector, a 
silicon strip detector and a transition radiation tracker, all 
immersed in a 2 T axial magnetic field. 
The calorimeter systems are of particular importance for the measurements presented in this paper.
The ATLAS calorimeter system  provides fine-grained measurements of shower energy depositions over a wide range of $\eta$. 
A highly segmented electromagnetic liquid argon (LAr) sampling
calorimeter covers the region  \mbox{$|\eta|<3.2$}, with granularity
that ranges from 0.003$\times$0.10 or 0.025$\times$0.025 to
0.1$\times$0.1 in $\Delta\eta\times\Delta\phi$, depending on depth
segment and pseudorapidity. It is divided into a barrel part ($|\eta|
< 1.475$) and an \ecap\ part ($1.375 < |\eta| < 3.2$). 
The hadronic barrel  \mbox{($|\eta|<1.7$)} calorimeter consists of steel absorbers and active scintillating tiles, with a granularity of either 0.1$\times$0.1 or 0.2$\times$0.1 depending on the layer.
The hadronic \ecap\ \mbox{($1.5<|\eta|<3.2$)} and forward  \mbox{($3.1<|\eta|<4.9$)} electromagnetic and hadronic calorimeters use liquid argon technology. The granularity in the hadronic \ecap\ ranges from 0.1$\times$0.1 to 0.2$\times$0.2. 
In the forward calorimeter, the cells are not arranged in projective towers but are  aligned parallel to the beam axis. As such the readout granularity is not constant in $\eta$--$\phi$.

Minimum bias trigger scintillator (MBTS) detectors are mounted in front of the \ecap\ calorimeters on both sides of the interaction point and cover the region \mbox{$2.1<|\eta|<3.8$}. The MBTS is divided into inner and outer rings, both of which have eight-fold segmentation, and is used to trigger the events analysed in this paper. 
\section{Event reconstruction}
\label{sec:reco}
This analysis is based on topological 
clusters in the calorimeter, which represent an attempt to reconstruct three-dimensional energy depositions associated with individual particles~\cite{cite:clustering}. The topological clustering algorithm proceeds through the following steps.
First, seed cells are found that have \mbox{$|E|>4\sigma$} above the 
noise level, where $E$ is the cell energy measured at the electromagnetic scale\footnote{
The electromagnetic scale is the basic calorimeter signal scale for the ATLAS calorimeters. It gives the 
correct response for the energy deposited in electromagnetic showers, but does not account for the lower 
response to hadrons.} and calibrated using test-beam data~\cite{cite:central-test-beam, cite:tile-test-beam, cite:lar-test-beam, cite:fcal-test-beam}. 
Next, neighbouring cells are collected into the cluster if they have \mbox{$|E|>2\sigma$} above the noise level. Finally, all surrounding cells are added to the cluster until no further cells with \mbox{$|E|>2\sigma$} are among the direct neighbours. 

The detector-level $\sumet$ is formed by summing the $\Et$ of all \topos\ in the
$\eta$--$\phi$ region of interest. 
Negative energy clusters are
included, leading to a convenient cancellation of the contributions from noise, which  can be either negative or positive.

To correct these clusters back to the particle-level, it is first necessary to determine the particle momenta to which the ATLAS calorimeters are sensitive.
Using a \geant~\cite{cite:geant} simulation of the ATLAS detector~\cite{cite:fullsim}, generator-level particles are propagated from the primary vertex to the calorimeters
 and the fraction of their energy deposited in the
calorimeters as \topos\ is studied as a function of $|\eta|$ and $p$
of the particle.
As discussed in section~\ref{sec:definition}, charged particles with $p >500$~MeV and neutral particles with \mbox{$p>200$ MeV} are found to deposit enough energy in the calorimeter to be included in the particle-level definition for all $|\eta|$ regions. 
Particles with lower momenta contribute a negligible
amount to the \topo\ $\sumet$ and are therefore excluded from the
particle-level $\sumet$ definition.

In order to properly correct for detector effects, the detector simulation must accurately describe the energy response of the calorimeters to low energy particles. The simulation calibration is refined using the di-photon invariant mass distribution of $\Pgg$ candidates. In data selected with the MBTS trigger, pairs of photon candidates in a given $\eta$ region are formed and their invariant mass, $\Mgg$, is constructed. In order to reduce combinatorial background, only events with exactly one pair in the $\eta$ region are considered.
 The data are compared to MC signal plus background templates in $\eta$ bins, which are chosen to reflect the boundaries of the calorimeter sub-systems.

\begin{figure}
	\begin{center}
	\subfigure[]{
		\includegraphics[scale=0.36]{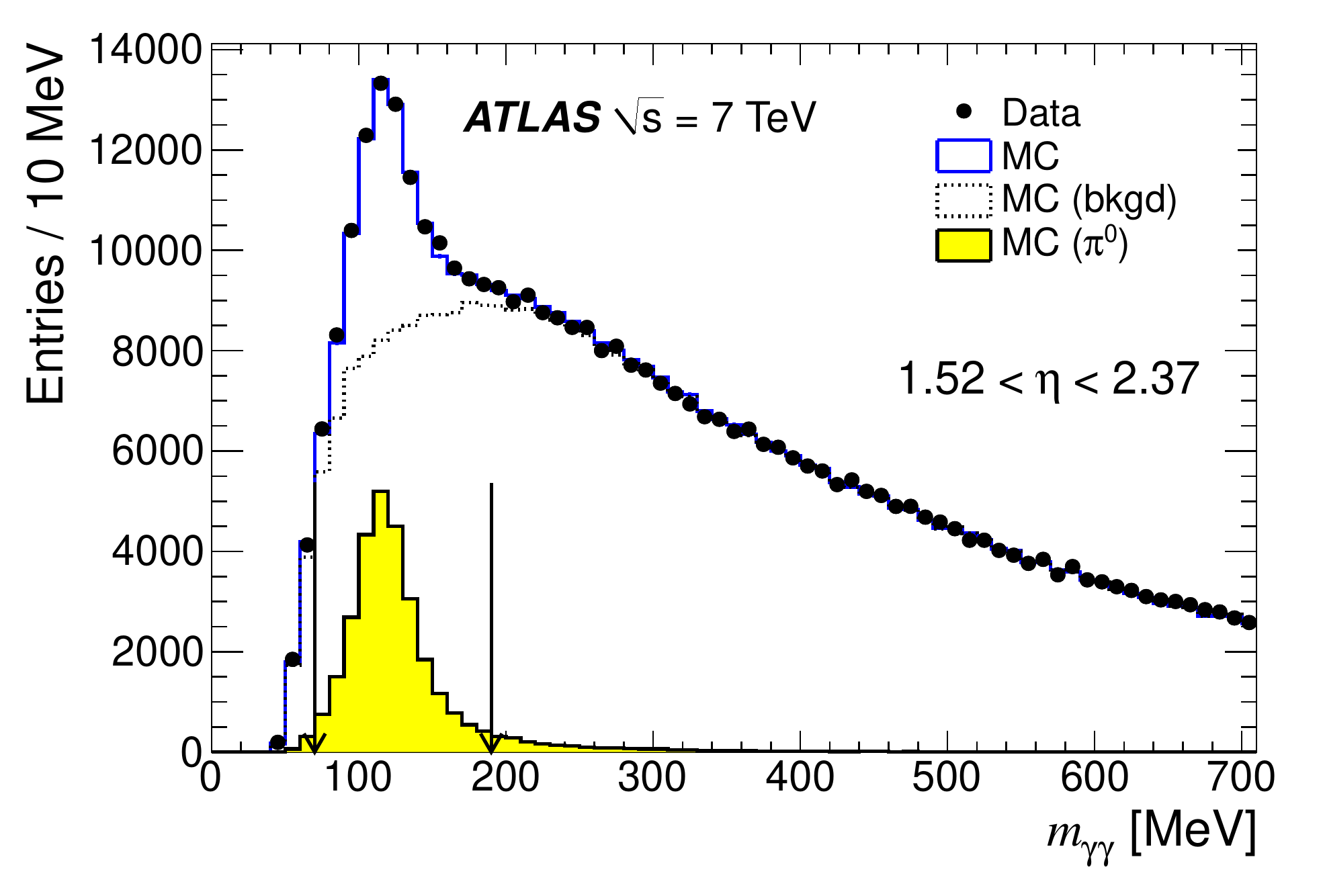}
	}
	\subfigure[]{
		\includegraphics[scale=0.36]{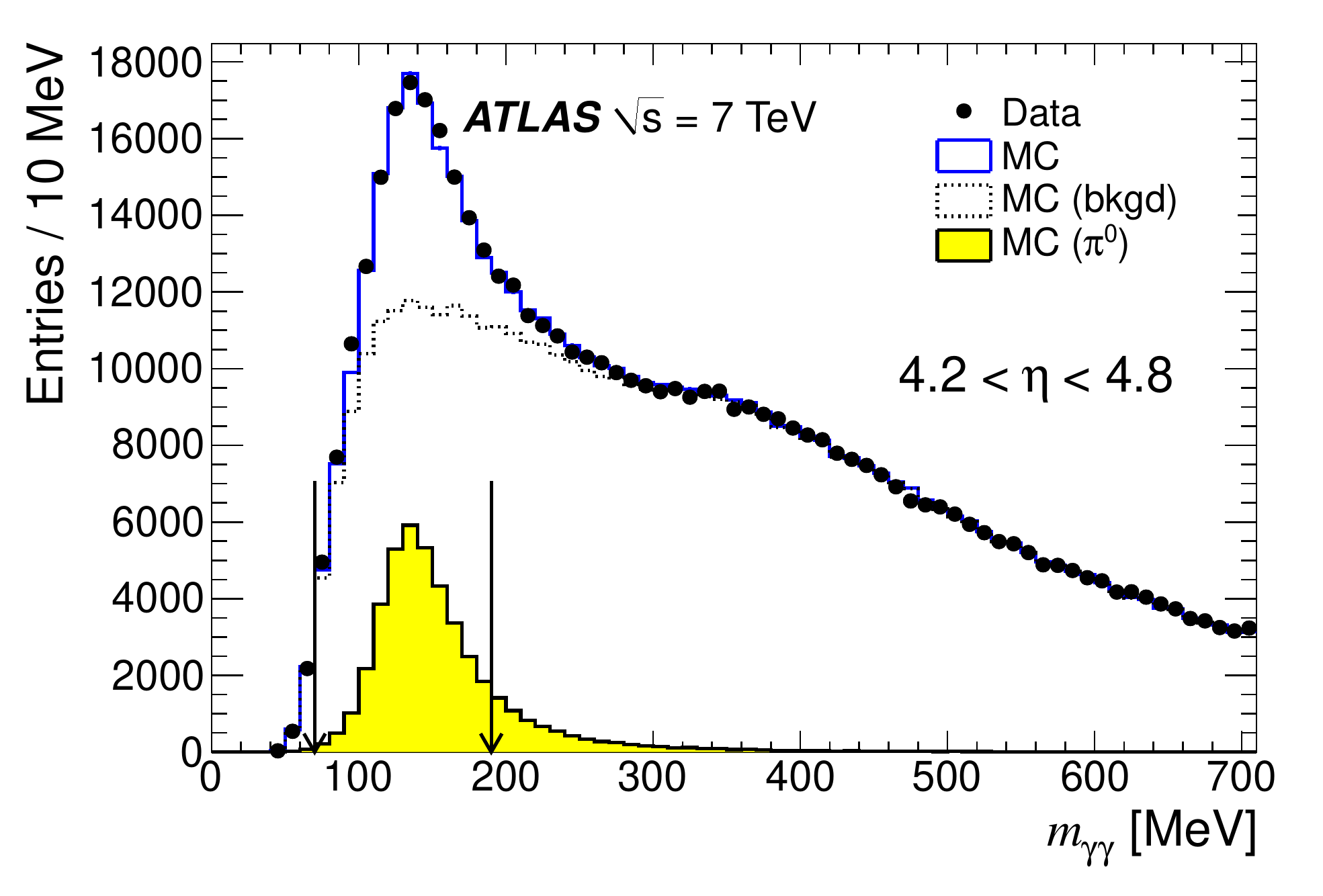}
	}
	\end{center}
	
	\caption{The di-photon invariant mass in the region (a) \mbox{$1.52<\eta<2.37$} and (b) \mbox{$4.2<\eta<4.8$}. The data are compared to the MC simulation with the best fit scale factor applied (this is 0.97$\pm$0.02 for (a) and 1.01$\pm$0.02 for (b)). The contribution from the MC signal $\Pgg$ templates and background templates are also shown separately. The arrows indicate the fit range.}
\label{fig:pi0}
\end{figure}

The signal
templates are derived from the \pythiasix\ \ambt\ samples, by matching pairs of \topos\ to generator level photons from a
$\pizero$ decay. The background templates are obtained using pairs of clusters that are not matched.
 The energies of the \topos\ in the signal template are
scaled by an energy response scale factor. 
This is varied and the $\chi^{2}$ between the data and MC distributions is minimized in order to determine the best fit value. Deviations from unity are typically 2--3\%\ but reach values of up to 10\%\ in some $\eta$ regions.
 This scale factor is then
applied to the energy of the MC \topos\ before unfolding the data.  
Figure~\ref{fig:pi0} shows the $\Mgg$ distribution in data compared to
the MC in two sample $|\eta|$ regions with the best fit scale factor
applied. 
\section{Event selection}
\label{sec:selection}
Events in the minimum bias analysis are selected with a one-sided MBTS trigger, which requires one counter on either side of the detector to be above noise threshold, suppressing contributions from empty beam crossings  and beam-induced background.
In order to suppress these contributions further, events  are required to have a reconstructed primary vertex with at least two
associated tracks with \mbox{$\Pt>150$~MeV} and \mbox{$|\eta|<2.5$}. 
Note that the track $\Pt$ cut is lower than the 250 MeV particle-level cut described in section~\ref{sec:truth-mb-selection}. 
This is because tracking and vertex reconstruction inefficiencies result in events with at least two 150~MeV reconstructed tracks having the same $\etflow$ as events with at least two 250~MeV charged particles, according to the MC models considered in this analysis.

Furthermore, events having more than one reconstructed vertex with five or more tracks are vetoed to suppress contributions from multiple proton-proton interactions. Five tracks are required on the additional vertices so that events with secondary vertices from decaying particles are not vetoed. 

Events in the dijet analysis are also selected with the one-sided MBTS trigger
and are required to pass the same event selection criteria as the minimum bias analysis.
 In addition, they are
required to contain two back-to-back jets passing the same kinematic selection criteria as the particle-level jets described in section~\ref{sec:truth-dijet-selection}.
\section{Corrections for detector effects}
\label{sec:corrections}
\begin{figure}
	\begin{center}
	\subfigure[]{
		\includegraphics[scale=0.36]{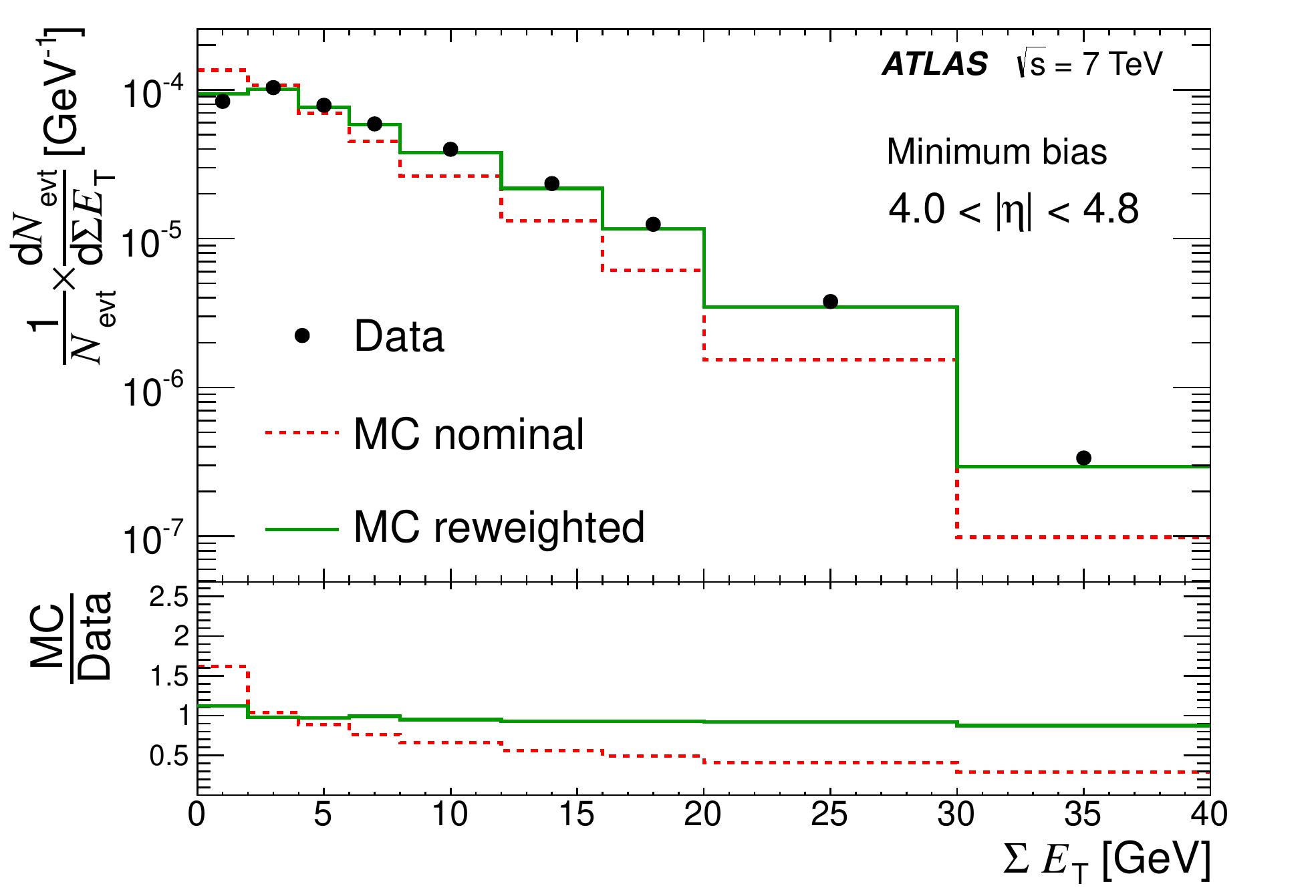}
	}
	\subfigure[]{
		\includegraphics[scale=0.36]{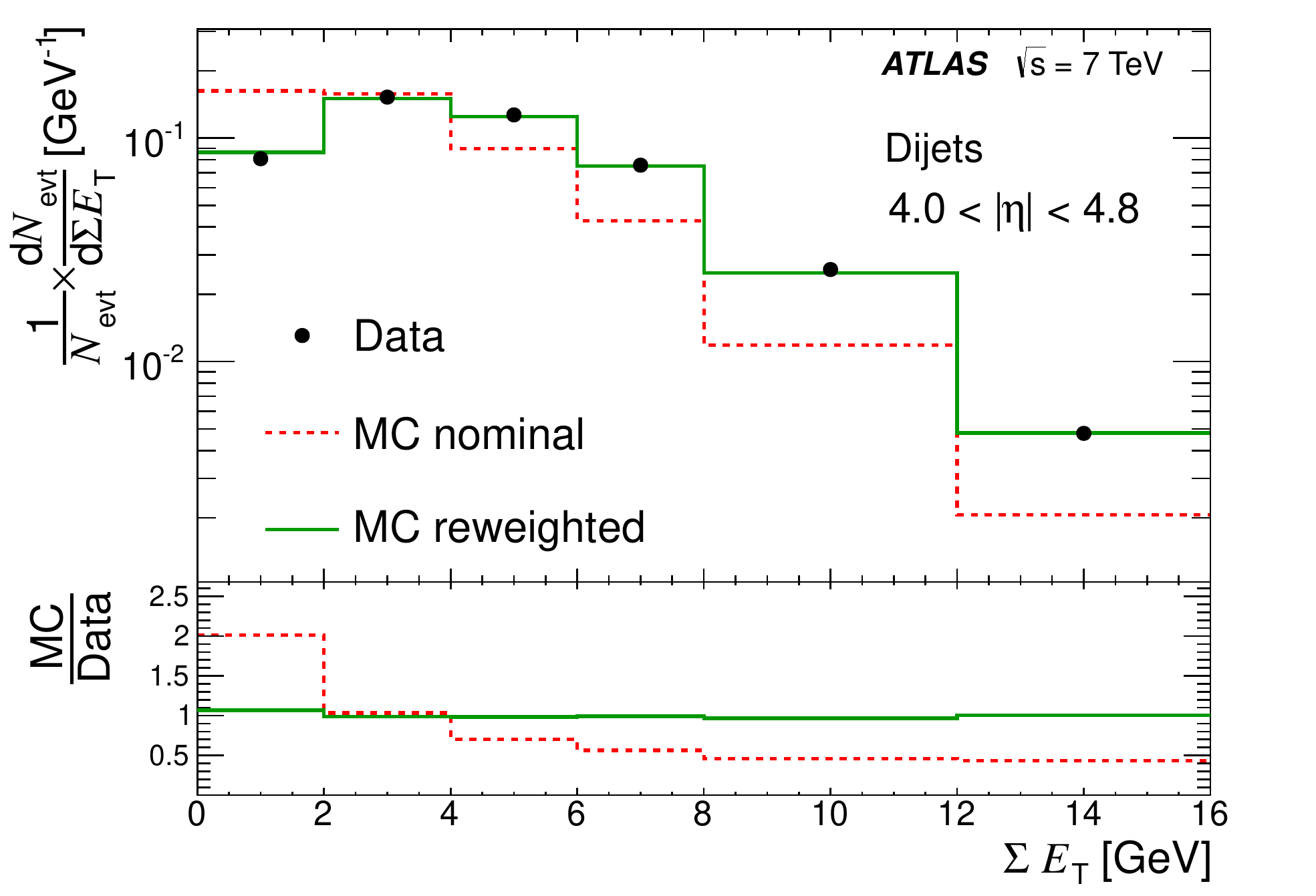}
	}
	\end{center}
	
	\caption{The detector-level $\sumet$ distribution in the region \mbox{$4.0<|\eta|<4.8$} for data compared to the nominal detector-level \pythiasix\ \ambt\ prediction and the reweighted detector-level \pythiasix\ \ambt\ prediction in (a) the minimum bias events and (b) the dijet events.}
\label{fig:reweight}
\end{figure}

The $\sumet$ distributions are unfolded in each $|\eta|$ region using an iterative Bayesian unfolding technique~\cite{cite:unfold}.
The $\etflow$ distribution is obtained by taking the mean of each unfolded $\sumet$ distribution and dividing by the $|\eta|$ and $\phi$ phase space. 
An unfolding matrix is formed from  events generated with \pythiasix\ \ambt, passed through the \geant\ simulation of the ATLAS detector.
The detector simulation accounts for energy losses of the
particles in material upstream of the calorimeter, for charged
particles that bend in the magnetic field and get swept out of the
calorimeter acceptance, and for the calorimeter response and resolution. 
Before unfolding each $\sumet$ distribution, the MC is reweighted by a fit to the ratio of the data to the MC detector-level $\sumet$ distribution, so that the $\sumet$ distribution matches that seen in data.
The MC significantly underestimates the $\sumet$ in the forward region, as seen in figure~\ref{fig:reweight}, where the detector-level $\sumet$ distribution in the region \mbox{$4.0<|\eta|<4.8$} is shown for both data and MC, before and after reweighting, for both the minimum bias and dijet selections.

The unfolding matrix associates the $\sumet$ formed from \topos\  with the $\sumet$ formed from generator-level particles. Events that pass the detector-level but not the particle-level selection criteria and vice versa are also accounted for in the correction procedure. The prior distribution of the particle-level $\sumet$ is initially taken from \pythiasix\ \ambt\ (reweighted to data) and the unfolding procedure is iterated twice, with the prior distribution replaced by the unfolded distribution after each iteration. A stable result is achieved after two iterations.
\section{Systematic uncertainties}
\label{sec:systs}
The dominant systematic uncertainties arise from three sources: (1) the accuracy with
which the MC simulates the energy response of the calorimeters to low
energy particles, (2) the knowledge of the amount of material upstream of the
calorimeters and (3) the MC generator model dependence in the
unfolding. In the dijet analysis an additional uncertainty arises from
the accuracy with which the MC simulates the jet energy scale. 
These sources are discussed in the following sub-sections. 
In each case the uncertainty on the unfolded data is obtained by shifting the MC by \mbox{$\pm 1\sigma$} for the source in question and comparing with the nominal unfolded data.
In order to give information about the correlations of the systematic uncertainties between
bins  and between the different distributions in this paper, each source 
is split into different components. These systematic uncertainties are summarized in tabular form in appendix~\ref{sec:tables}.

The following additional potential sources of systematic uncertainty are 
found to be negligible: energy resolution, multiple proton-proton interactions, contributions from
noise and beam-induced backgrounds, simulation of the primary vertex position, simulation of the trigger selection, and simulation of the position of the forward calorimeter. 
\subsection{Calorimeter energy response}
The systematic uncertainty on the calorimeter energy response is determined separately for electromagnetic and hadronic particles. An average is then obtained, using the \pythiasix\ \ambt\ prediction of the relative contributions to the  $\sumet$ by different particle types.
For electromagnetic particles the systematic error comes from
 uncertainties on the extraction of the energy scale from fits to the  $\Mgg$ distributions in $\Pgg$ candidates. These
are obtained from variations in the fit range, the background shape, the
criteria for matching reconstructed photons to generator-level photons in the production of the signal template, variations
in the simulation of the calorimeter resolution, and consistency with a
similar analysis using tighter kinematic and photon identification cuts. 
The total uncertainty depends on the $|\eta|$ region and is generally at the level of 2--4\%, 
but increases up to 15\%\ in the regions where different calorimeter sub-systems overlap.

The uncertainty on the energy response for hadronic particles in the
central region, where there is good coverage from the inner tracking detector, is obtained from studies of the ratio of the calorimeter energy
measurement to the inner detector track momentum measurement, for
isolated charged pions~\cite{Aad:2012vm}.  
The uncertainty is obtained by taking the difference between data and MC in $p$ and $|\eta|$ bins and is found to be 
3.5\%\ for \mbox{$|\eta|<0.8$} and 5\%\ for \mbox{$0.8<|\eta|<2.4$}.
In the forward region the energy response
uncertainty for hadrons is taken from the difference between the MC and data in test-beam studies of charged pions~\cite{cite:forward-test-beam}.
This leads to a one-sided uncertainty for hadrons relative to electromagnetic particles of $+5\%$ in the region  \mbox{$2.5<|\eta|<3.2$} and $+9\%$ in the forward calorimeter \mbox{($|\eta| > 3.2$)}. 

The only component of the systematic uncertainty on the energy response assumed to be correlated between $|\eta|$
bins is that of the forward calorimeter (determined from test-beam results), which affects
the bins  \mbox{$3.2<|\eta|<4.0$} and \mbox{$4.0<|\eta|<4.8$}. The systematic uncertainties from the $\Pgg$ fits are assumed to be uncorrelated as the $\Mgg$ shapes are rather different in the different $|\eta|$ regions, resulting in different possible systematic shifts. Similarly, the difference between data and MC for the ratio of calorimeter energy to inner detector track momentum does not show systematic shifts in one direction and is assumed to be uncorrelated. 

Appendix~\ref{sec:tables} gives both the uncorrelated and correlated uncertainties in each bin of each distribution. The former vary between 2.4\%\ and 5.4\%\ for the $\etflow$ in the minimum bias data, depending on the $|\eta|$ region. 
The largest uncertainty is in the region \mbox{$0.8<|\eta|<1.6$},
which contains the region of overlap between the barrel and \ecap\
electromagnetic calorimeters ($1.375 < |\eta| < 1.475$).
The correlated source is about $-6\%$ for the two highest $|\eta|$ bins in the minimum bias data and about $-8\%$ in the dijet data. Note that a positive uncertainty on the energy scale in the MC leads to a negative uncertainty on the corrected result in the data.
The uncertainty is higher in the dijet data, due to a larger contribution from events where the detector-level jets pass the selection criteria but the generator-level jets do not. 
Their $\sumet$ distribution is taken from the MC so a shift in the energy scale leads to an additional bias in the corrected result. 
\subsection{Material description}
The amount of material upstream of the calorimeters affects the $\sumet$ distributions because particles can interact and lose some of their energy before reaching the calorimeter. It is therefore important to have a realistic description of the material in the MC simulation used to perform the detector corrections. 

In order to assess the systematic uncertainty arising from possible discrepancies in the material description, detector corrections are recalculated using a special \pythiasix\ \ambt\ sample with additional material. The sample is based on a similar one described in section 3 of ref.~\cite{Aad:2011mk}, but with additional material introduced in the forward region. The results are compared to the nominal unfolded data and the difference is taken as a symmetric systematic uncertainty to account
 for the possibility of the MC simulation either underestimating or overestimating the amount of material.

In order to understand the correlations between the uncertainties in different $|\eta|$ bins, the additional material is split into three components:  
(1) extra material upstream of the barrel calorimeter, (2) an increase in material density in the barrel-\ecap\ overlap region and (3) additional material in the inner detector, the inner detector services and the  forward region, as well as an increase in the material density in some detector volumes in the forward region.
The systematic uncertainties arising from components (1) and (2) are assumed to be correlated between  $|\eta|$ bins, whereas the uncertainty arising from component (3) is assumed to be uncorrelated, due to the fine structure of these detectors with respect to the wide bins used in this analysis. 

Source (1) affects only the first two $|\eta|$ bins, at the level of about 3\%\ in the minimum bias data and 1.3--2.5\%\ in the dijet data. The uncertainty is generally smaller in the dijet data as the particles in these events tend to have larger momenta.
 Source (2) affects only the second and third bins and is less than 1\%. Source (3) affects all $|\eta|$ bins and ranges between 0.23\%\ and 5.5\%, with the largest uncertainty in the region \mbox{$1.6<|\eta|<2.4$}, where there is a large amount of material associated with the inner detector.

\subsection{Physics model dependence}
\begin{figure}
	\begin{center}
	\subfigure[]{
		\includegraphics[scale=0.36]{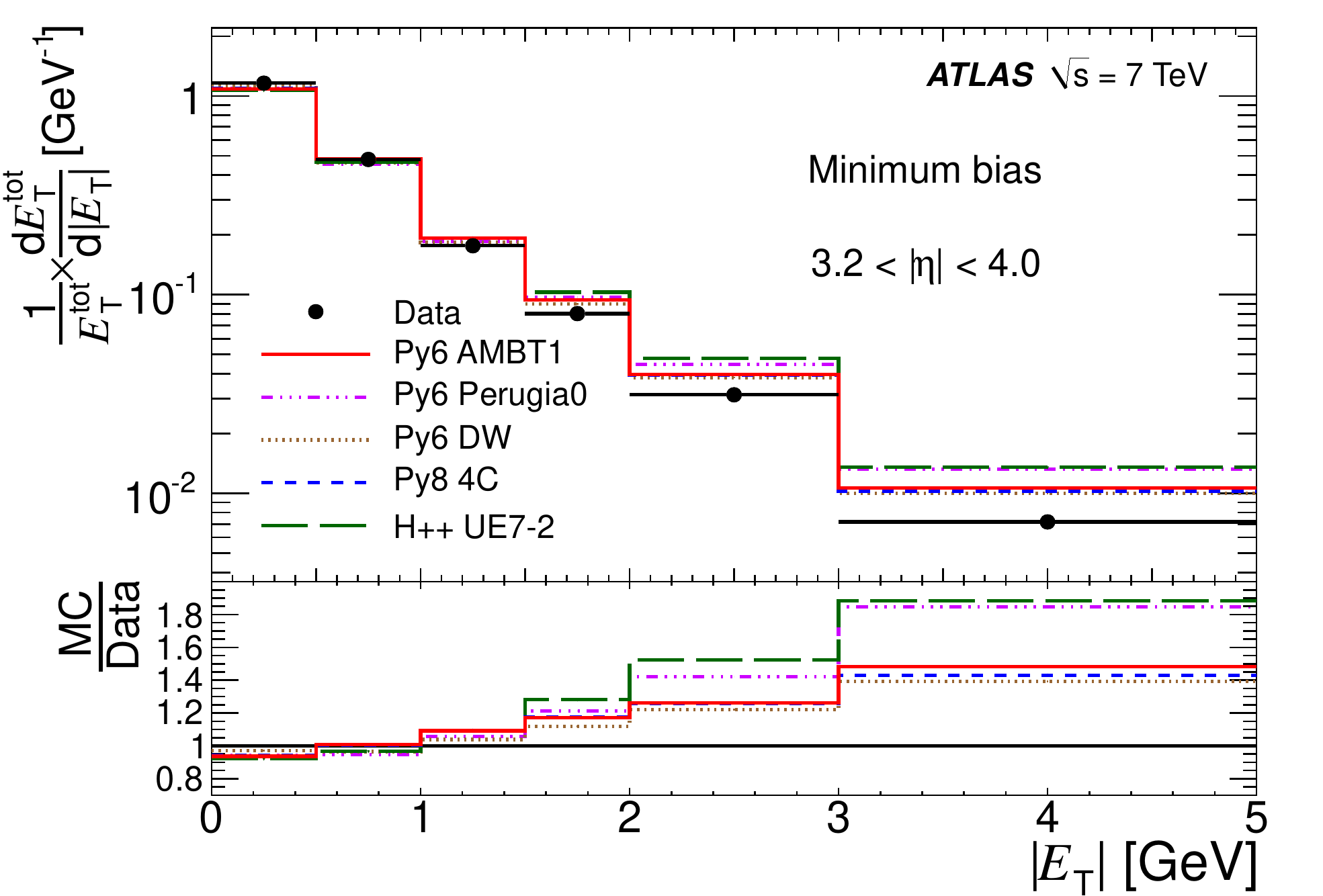}
                \label{fig:et_mb}
         }
	\subfigure[]{
		\includegraphics[scale=0.36]{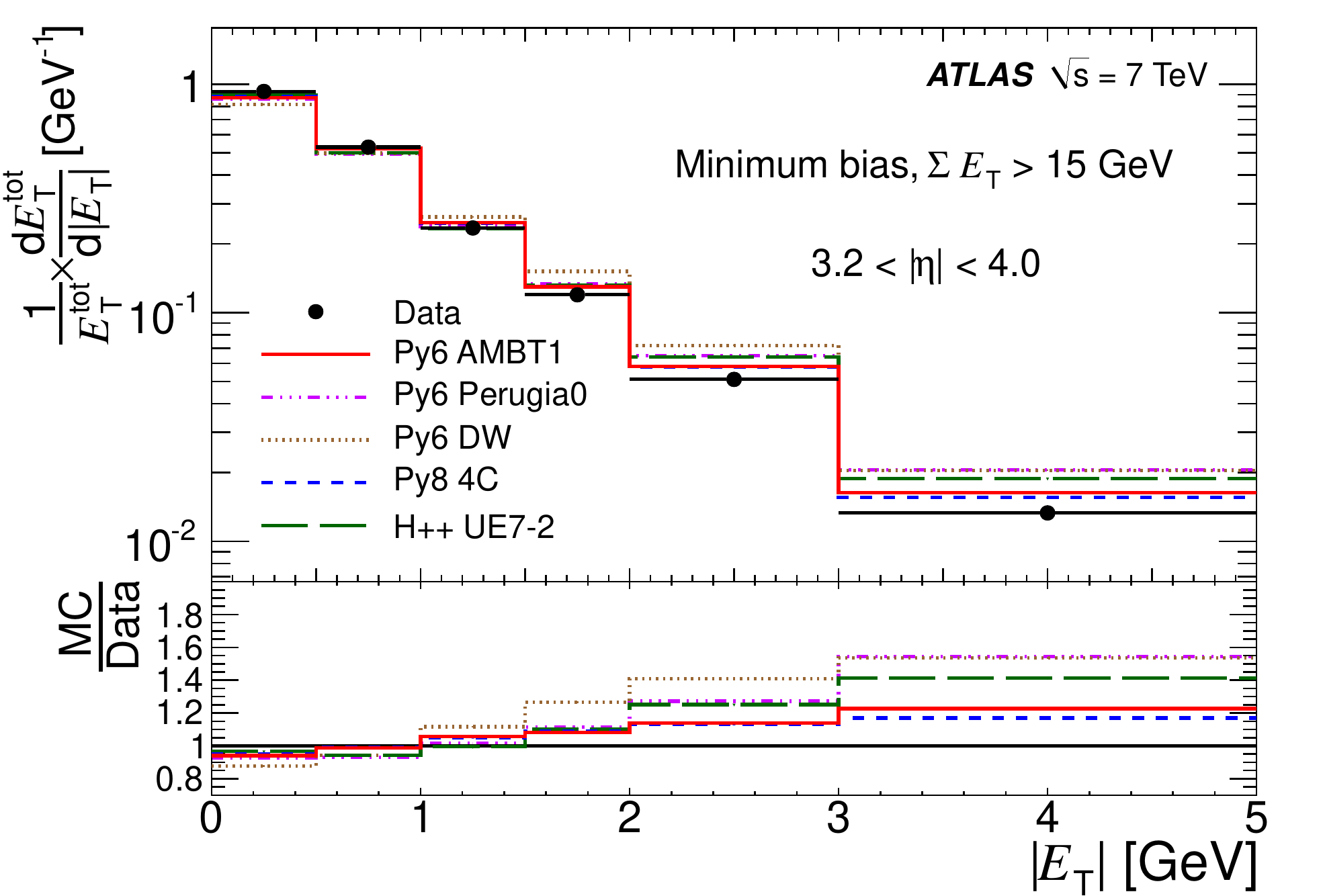}
                \label{fig:et_mb_highsumet}
	}
	\subfigure[]{
		\includegraphics[scale=0.36]{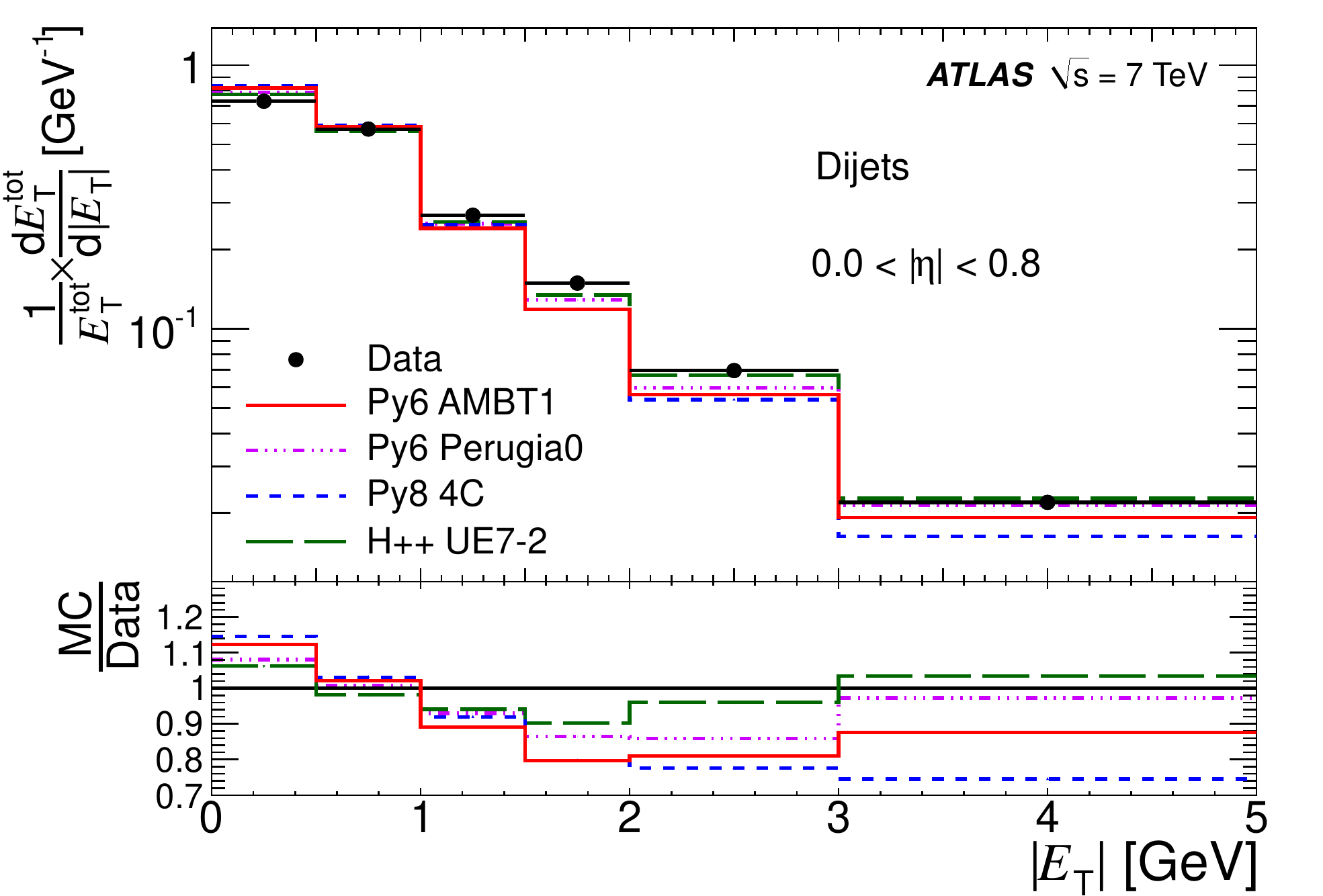}
                \label{fig:et_dj}
	}
	\subfigure[]{
		\includegraphics[scale=0.36]{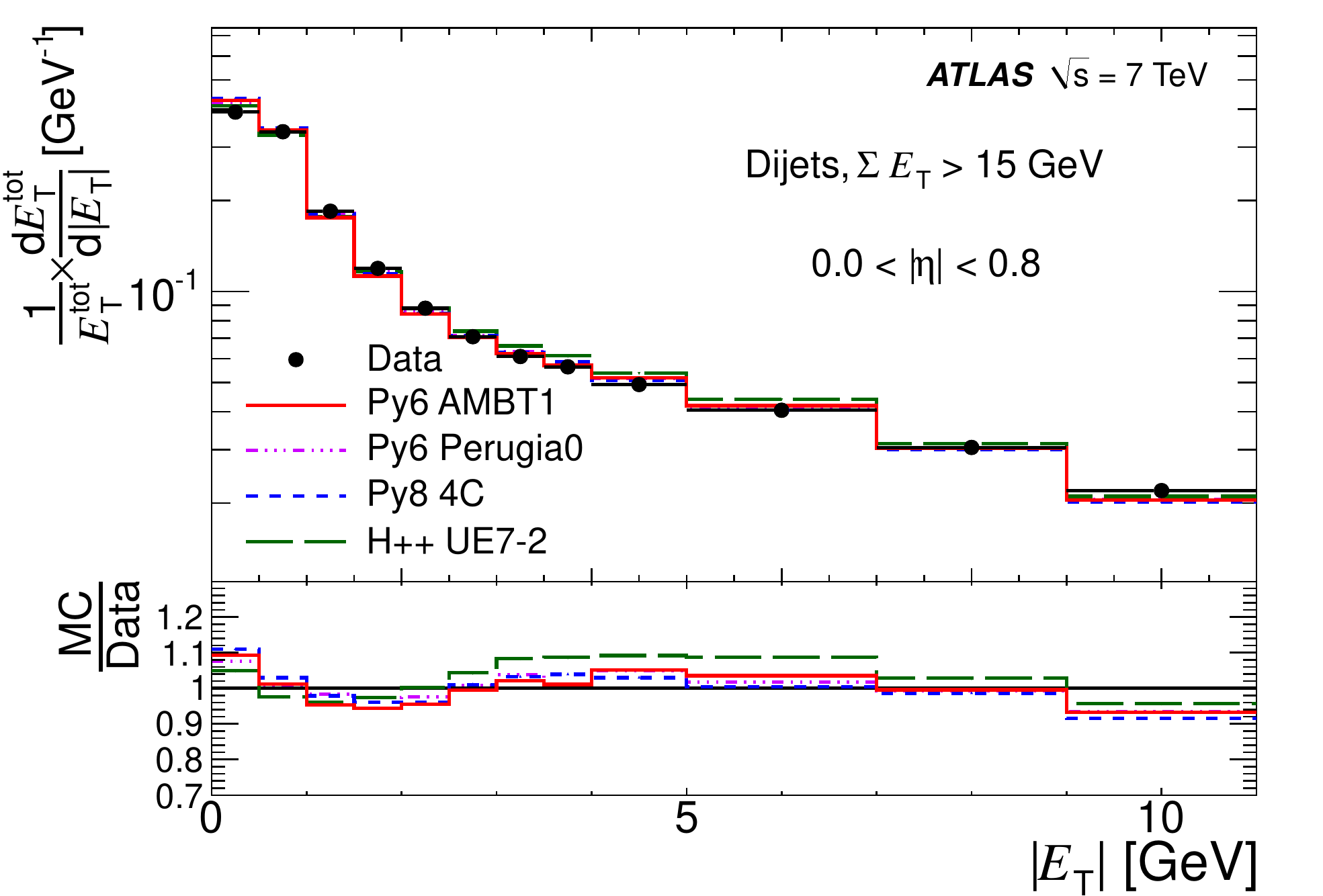}
                \label{fig:et_dj_highsumet}
}
	\end{center}	
	\caption{Distribution of $\sumettotlong$, where $\sumettot$ is the sum over events of the detector-level $\sumet$, and $\Et$ is the detector-level cluster transverse energy (a) in minimum bias events in the region \mbox{$3.2<|\eta|<4.0$}; (b) as in (a) but for events with \mbox{$\sumet > 15$ GeV}; (c) in dijet events in the region \mbox{$0.0<|\eta|<0.8$}; (d) as in (c) but for events with \mbox{$\sumet > 15$~GeV}. The data are compared to various MC predictions.}
\label{fig:et}
\end{figure}

The MC model used to correct the data can affect the results as a realistic description of particle kinematics is needed. The model dependence is minimized by first reweighting the detector-level MC to the data and then by iterating the unfolding, using the unfolded data as the new prior distribution after each iteration. This reduces the dependence on the $\sumet$ spectrum itself; however, other kinematic distributions can also affect the unfolding. One important variable is the $\Et$ of the individual particles, as the calorimeter response to a particle is energy dependent. 
The dependence on the model is investigated by performing the unfolding with other MC models.
The following  MC models and tunes are considered: \pythiasix\ \ambt\ (nominal), \pythiasix\ \DW, \pythiasix\ \perugia, \pythiaeight\ \fourc\ and \hpp\ \ueseven. Details of these tunes are given in table~\ref{tab:tunes}.
The MC model used to assess the systematic uncertainty is chosen to ensure a reasonable spread in the particle kinematics with respect to the reference \pythiasix\ \ambt\ model.
Figure~\ref{fig:et} shows distributions of $\sumettotlong$, where $\sumettot$ is the sum over events of the detector-level $\sumet$, and $\Et$ is the detector-level cluster transverse energy.
These distributions show the relative contribution to the $\sumet$ from clusters with a given $\Et$.
$|\Et|$ is plotted instead of $\Et$ since the former leads to a cancellation in the contribution from noise.
Figure~\ref{fig:et_mb} shows the distribution in minimum bias events for the region \mbox{$3.2<|\eta|<4.0$}. This region is shown as it has significant differences between data and MC.
The contribution to the $\sumet$ from high $\Et$ clusters is smaller in data than in \pythiasix\ \ambt. The model with the largest deviations from \pythiasix\ \ambt\ is \hpp\ \ueseven, indicating that this model can be used to assess possible biases in the unfolding due to this effect. 
It should be noted that at high $|\Et|$ \hpp\ \ueseven\ lies above \pythiasix\ \ambt\, while the data lie below it, but the final systematic uncertainty is symmetrized. 
The same distribution is shown in figure~\ref{fig:et_mb_highsumet} for the sub-sample of events with \mbox{$\sumet > 15$~GeV}. Again, the data have a softer cluster $|\Et|$ distribution. Here  \pythiasix\ \DW\ shows the largest deviations from \pythiasix\ \ambt. Since unfolding with \pythiasix\ \DW\ results in a larger shift in the corrected data than unfolding with \hpp\ \ueseven, the former is used to assess the systematic uncertainty in the minimum bias events.

Figure~\ref{fig:et_dj} shows the same distribution in dijet events; the most central region is shown as the differences between data and MC are largest in this region. This time the data distribution is harder than \pythiasix\ \ambt. Again \hpp\ \ueseven\ has the largest deviations. Figure~\ref{fig:et_dj_highsumet} shows the same distribution for events with \mbox{$\sumet > 15$~GeV};  all the models agree well with the data, but \hpp\ \ueseven\ has the largest deviations. For the dijet selection \hpp\ \ueseven\ is therefore used to assess the systematic uncertainty.

For both the minimum bias and dijet analyses, this systematic uncertainty is symmetrized and treated as correlated between  $|\eta|$ bins (although not correlated between the two analyses). 
The uncertainties on the $\etflow$ range from 2--4\%\ for the minimum bias data and are 2\%\ or less for the dijet data.

\subsection{Jet energy scale}
In the dijet selection, events are required to contain at least two jets with \mbox{$\Et > $ 20~GeV}.  It is possible that events that 
satisfy the detector-level criteria do not satisfy the criteria at the particle-level, and vice versa. This is accounted for in the correction procedure, but if there are differences in the jet energy scale between data and MC simulation this could result in a bias in the correction procedure. 
The uncertainty on the jet energy scale is described in ref.~\cite{Aad:2011he}. 
The corresponding uncertainty on the $\etflow$ is at the level of 1.6\%\ in the most central bin and decreases to 0.13\%\ in the most forward bin. It is treated as correlated between $|\eta|$ bins. For the $\sumet$ distributions this source of uncertainty is negligible and therefore neglected in the region $|\eta| > 2.4$.
\section{Results}
\label{sec:results}
\subsection{Nominal Results}
\begin{figure}
	\begin{center}     
          \subfigure[]{     
            \includegraphics[scale=0.58]{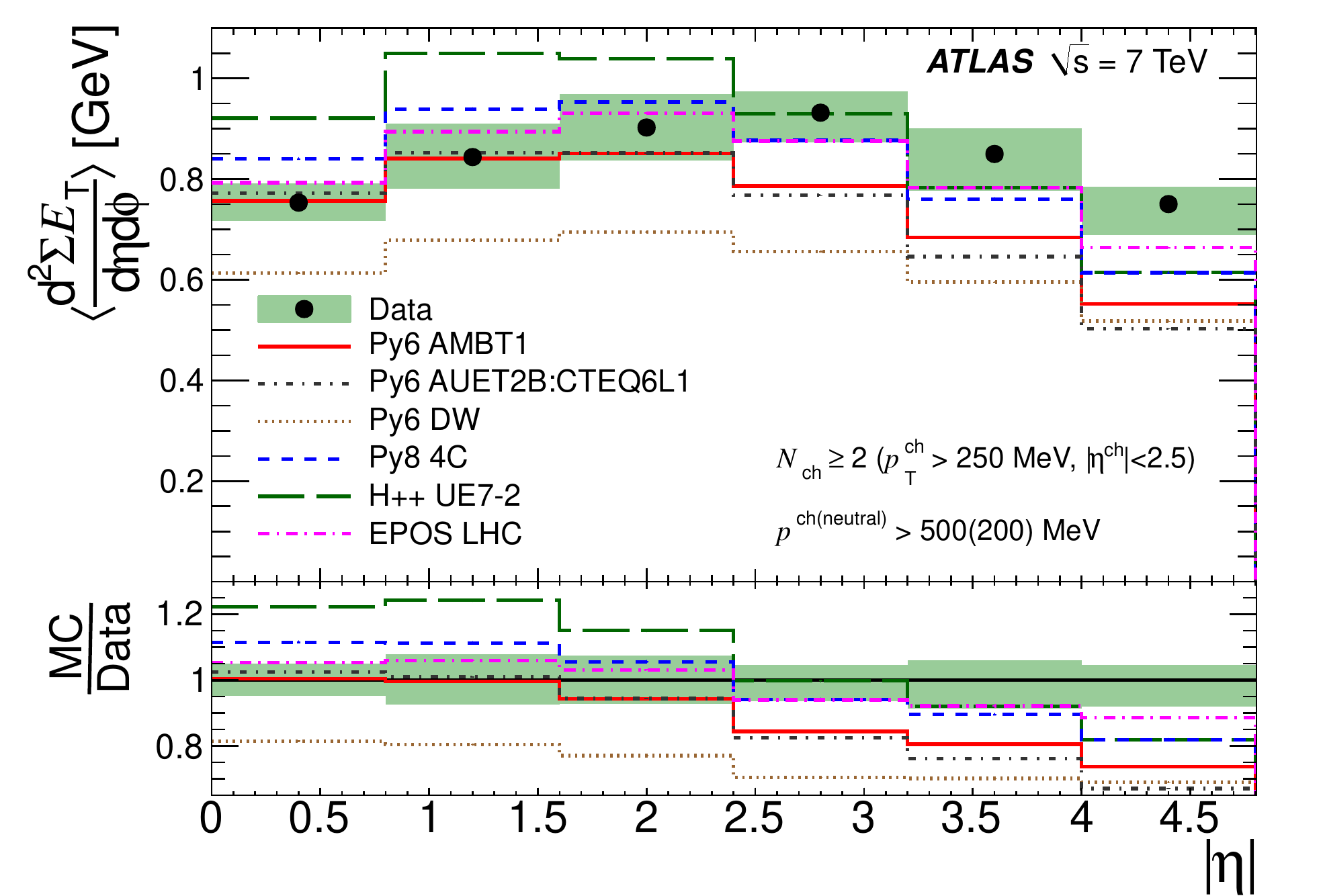}
	\label{fig:finalplot-mb}
          }
          \subfigure[]{     
            \includegraphics[scale=0.58]{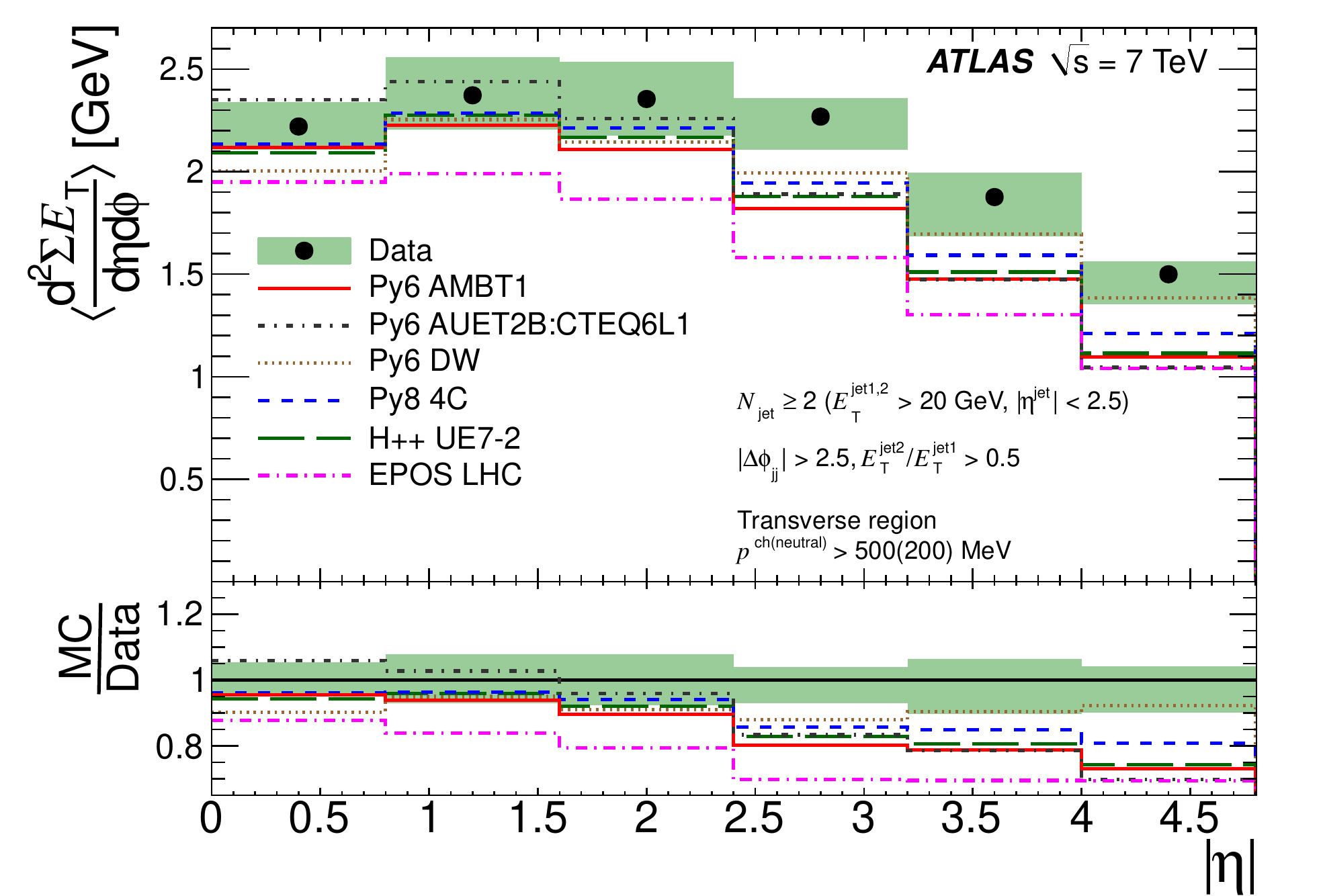}
	\label{fig:finalplot-dijet}
          }
	\end{center}
	\caption{Unfolded $\etflow$ distribution compared to various MC models and tunes for (a) the minimum bias selection and (b) the dijet selection in the transverse region. The filled band represents the total uncertainty on the unfolded data.
\vardefmb\
\vardefdj\
\vardef\
}
	\label{fig:finalplot}
\end{figure}

The unfolded $\etflow$ distributions are shown in figure~\ref{fig:finalplot} for both the minimum bias and the dijet selections.  The  filled bands indicate the systematic and statistical uncertainties on the data, added in quadrature. 
 In all bins the systematic uncertainty is significantly larger than the statistical uncertainty.   
The $\etflow$ distribution in the minimum bias data dips in the
central region. Since the relative fraction of low momentum particles
is higher in the central region than in the forward region, fewer
central particles pass the selection criteria described in
section~\ref{sec:definition}, hence reducing the $\sumet$ in the
central region. 
The dip in the central region is less prominent in the dijet data; this feature is discussed below.

\begin{figure}
	\begin{center}     
    
            \includegraphics[scale=0.55]{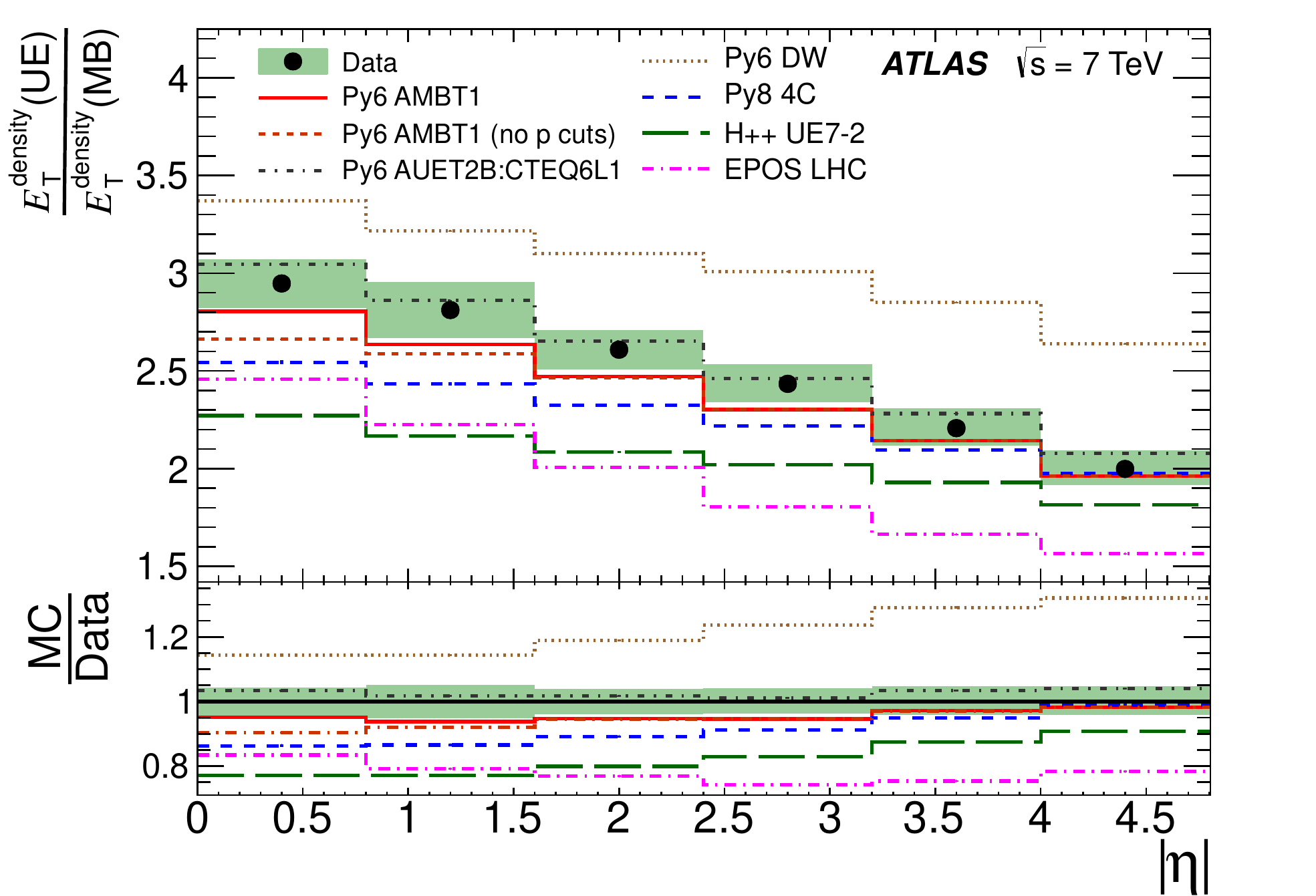}

	\end{center}
	\caption{Unfolded $\etflow$ distribution in the dijet data transverse region divided by that in the minimum bias data, compared to various MC models and tunes. The filled band represents the total uncertainty on the unfolded data.
}
	\label{fig:finalplotratio}
\end{figure}

Figure~\ref{fig:finalplotratio} shows the ratio of the $\etflow$ in the dijet transverse region to the $\etflow$ in minimum bias events.
The correlations between the systematic uncertainties for the dijet and minimum bias distributions are taken into account. All systematic uncertainties but the physics model dependence and jet energy scale are taken as correlated between the two.
The $\etflow$ in the transverse region for the dijet selection is larger than the $\etflow$ in the minimum bias data.
This increase is expected, due primarily to the presence of a hard scatter, which will bias the selected events away from peripheral proton scatters and towards small impact parameter (``head-on'') proton-proton interactions. This means that more parton-parton interactions are likely to occur in the underlying event in the dijet data than in the collisions with a larger impact parameter that characterize the events in the minimum bias dataset.

The unfolded data are compared to various MC models.
In the minimum bias sample the $\etflow$ distribution in figure~\ref{fig:finalplot-mb} is well described by \pythiasix\ \ambt\ in the central region. This is expected as this tune was prepared with ATLAS 7~TeV minimum bias data in the region \mbox{$|\eta|<$2.5~\cite{cite:mb}}.   
At higher $|\eta|$ values, however, the $\etflow$ is underestimated and is approximately $25\%$ too low in the highest $|\eta|$ bin. 
The \pythiasix\ \auetbtwo\ prediction is very similar to that from \pythiasix\ \ambt, with slightly more energy in the central region and less in the forward region, meaning that the description of the $|\eta|$ dependence is even worse.
\pythiasix\ \DW\ underestimates the $\etflow$ in all $|\eta|$ bins. 
Despite this it provides an improved description of the $|\eta|$ dependence of the $\etflow$. \pythiaeight\ \fourc\ overestimates the $\etflow$ in the central region. The agreement improves in the region \mbox{$1.6<|\eta|<3.2$}, but in the higher $|\eta|$ bins the $\etflow$ is underestimated. \hpp\  \ueseven\ overestimates the $\etflow$ in the central region, describes the data well in the region \mbox{$2.4<|\eta|<3.2$}, and undershoots the data at higher $|\eta|$. 
The \eposlhc\ prediction provides the best description over the entire $|\eta|$ region, although it does fall slightly too fast with $|\eta|$.
It should be noted that, with the exception of \eposlhc\ and \pythiasix\ \DW, while some models and tunes appear to agree better in some regions than others, this is generally due to differences in the total level of particle production. The overall pattern remains the same: the $\etflow$ in the forward region is too low relative to the central region.

In the dijet selection in figure~\ref{fig:finalplot-dijet}, all of the MC models and tunes perform reasonably well in the central region, apart from \eposlhc\ which underestimates the $\etflow$ in all $|\eta|$ bins.
\pythiasix\ \auetbtwo\ slightly overestimates the energy in the most central bins, and all the other predictions are slightly too low.
As was the case in the minimum bias analysis, the $\etflow$ in the forward region is underestimated.  \pythiaeight\ \fourc\ is approximately 20\%\ too low in the most forward bin, while \pythiasix\ \ambt, \hpp\  \ueseven\  and \pythiasix\ \auetbtwo\ are 25--30\%\ too low.  \pythiasix\ \DW\ provides the best description of the $|\eta|$ dependence, although the overall amount of energy is too low.

The fall-off with $|\eta|$ of the ratio of the $\etflow$ in dijet and minimum bias events seen in figure~\ref{fig:finalplotratio} is reproduced by the models, with \pythiasix\ \ambt\ and \auetbtwo\ describing the data the best. The reduction in the ratio with $|\eta|$ is partly due to the momentum cuts on the particles included in the $\sumet$ calculation. In the dijet data, the particles tend to have larger momenta and so fewer are removed from the $\sumet$ calculation. 
According to \pythiasix\ \ambt, the momentum cuts remove 25(18)\%\ of the $\etflow$ in the most central bin and a negligble amount in the most forward bin for the minimum bias (dijet) selections.
The \pythiasix\ \ambtnopcuts\ curve in figure~\ref{fig:finalplotratio},  shows the ratio when the momentum cuts on the particles contributing to the $\sumet$ have been removed. There is still a residual decrease with $|\eta|$ which may be due to a contribution to the underlying event in the central region coming from particles associated with the hard scatter.

\begin{figure}[]
	\begin{center}
\subfigure[]{
  \includegraphics[scale=0.36]{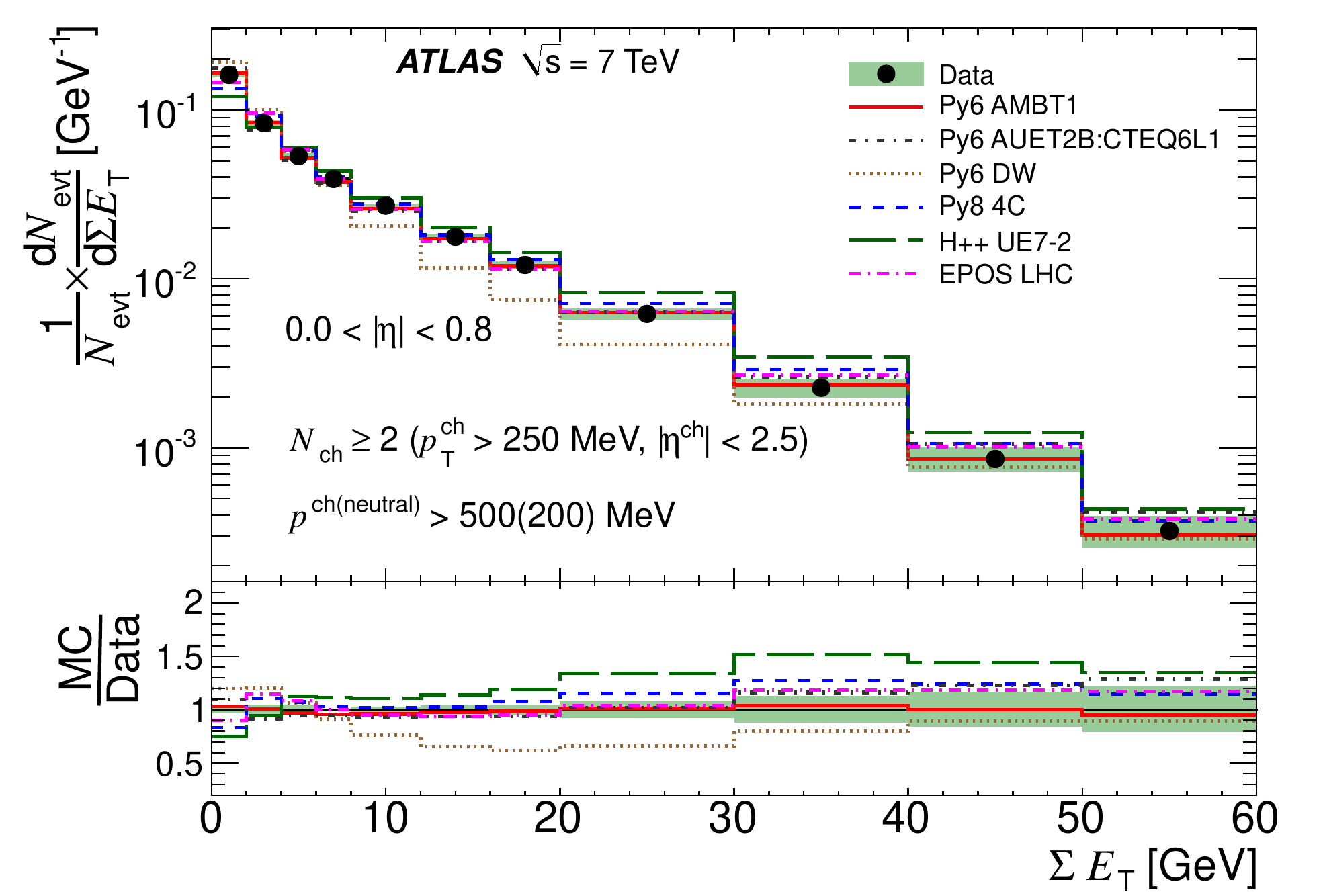}
}
\subfigure[]{
  \includegraphics[scale=0.36]{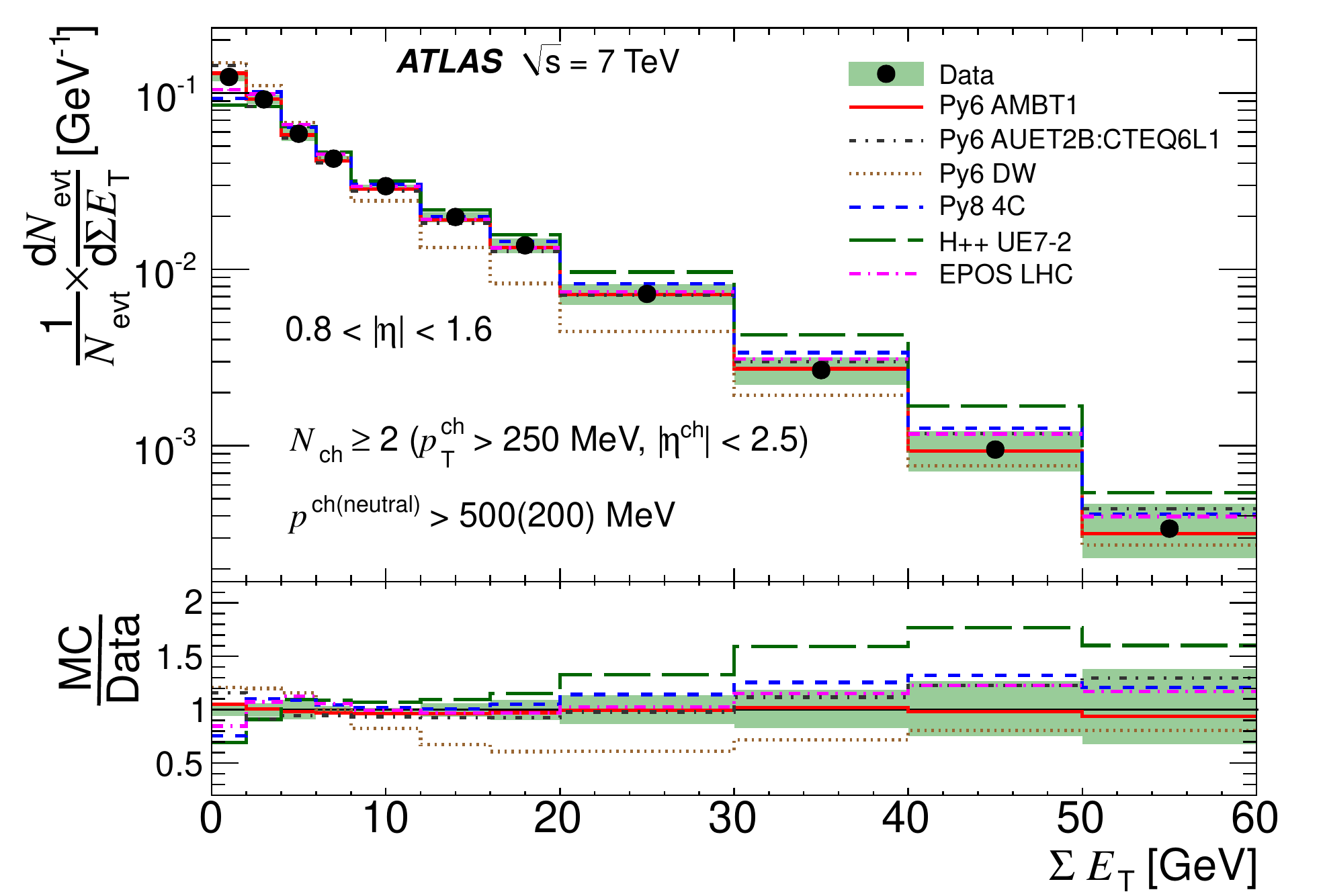}
}
\subfigure[]{
  \includegraphics[scale=0.36]{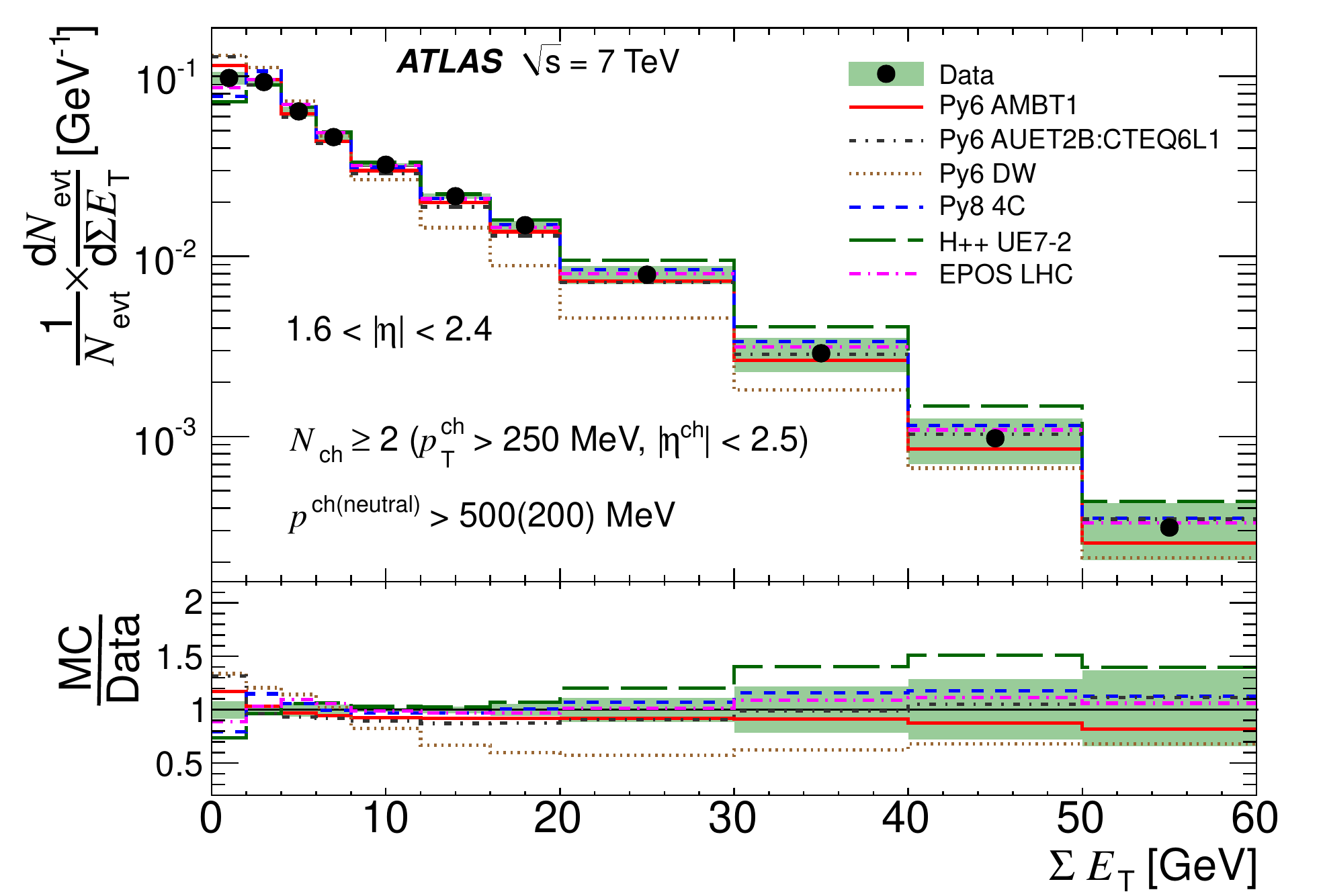}
}
\subfigure[]{
  \includegraphics[scale=0.36]{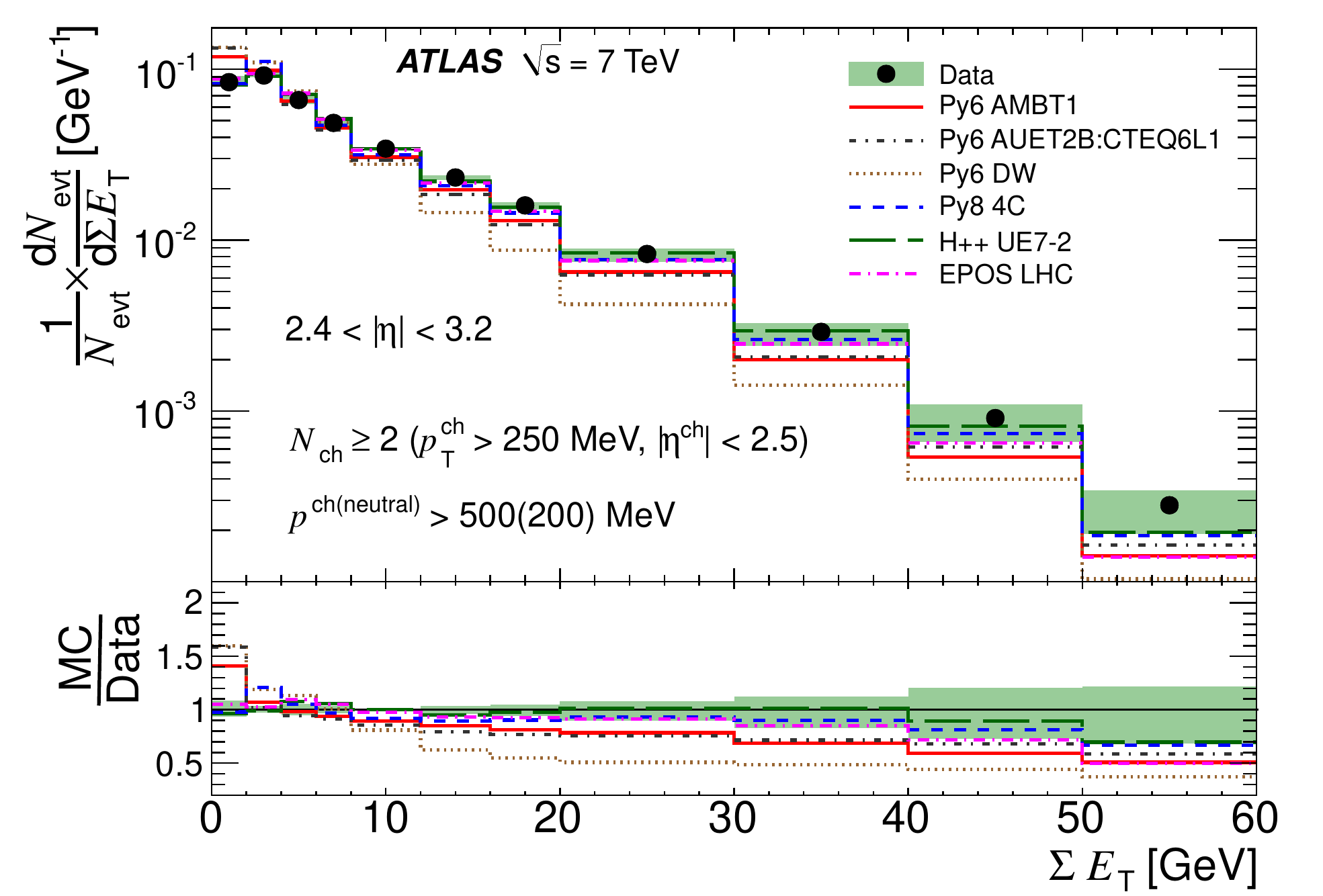}
}
\subfigure[]{
  \includegraphics[scale=0.36]{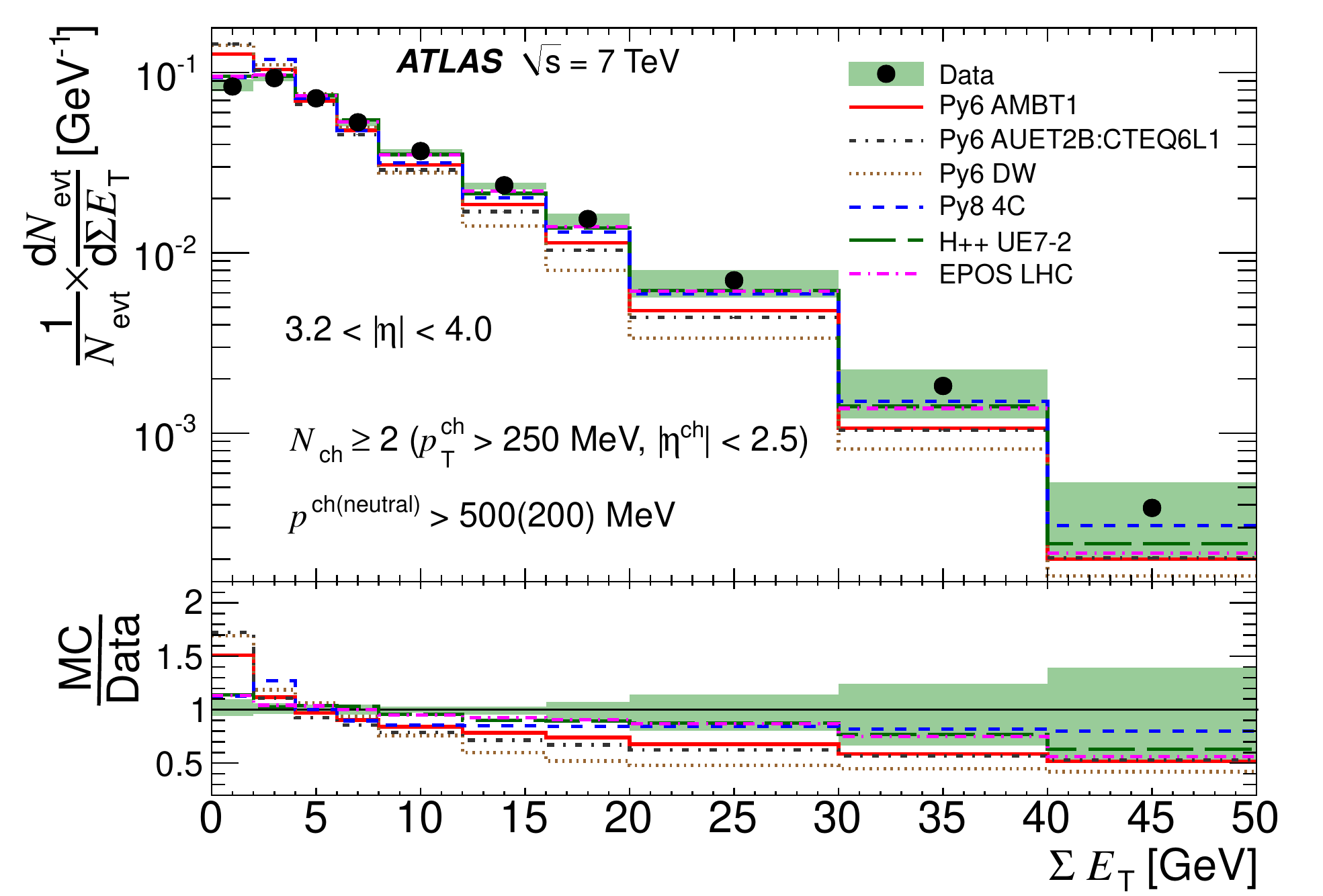}
}
\subfigure[]{
  \includegraphics[scale=0.36]{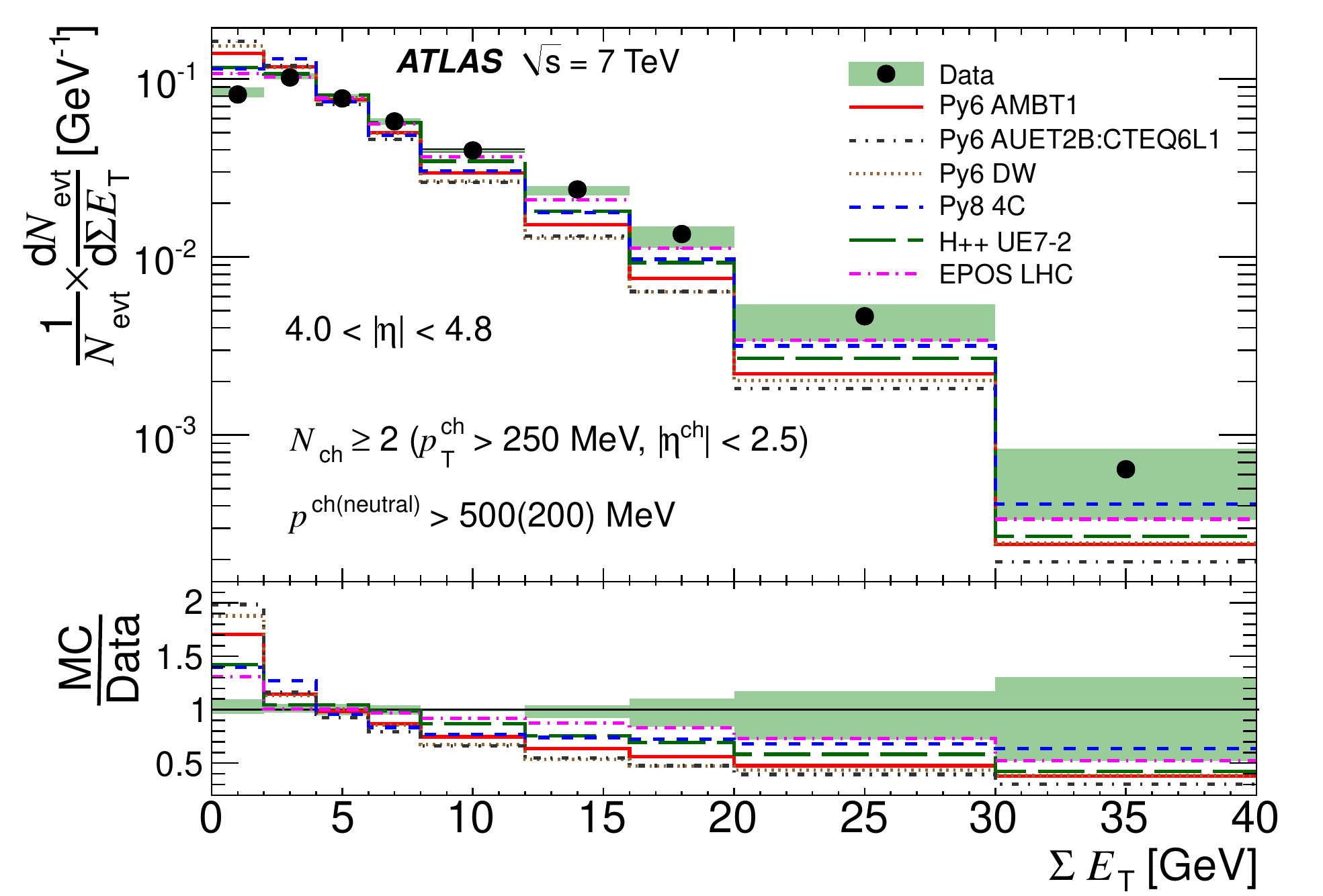}
}
	\end{center}
	\caption{Unfolded $\sumet$ distributions compared to various MC models and tunes for the minimum bias selection in the following $|\eta|$ regions:
(a) \mbox{$0.0<|\eta|<0.8$},
(b) \mbox{$0.8<|\eta|<1.6$},
(c) \mbox{$1.6<|\eta|<2.4$}, 
(d) \mbox{$2.4<|\eta|<3.2$}, 
(e) \mbox{$3.2<|\eta|<4.0$} and 
(f) \mbox{$4.0<|\eta|<4.8$}. 
The filled band in each plot represents the total uncertainty on the unfolded data. 
\vardefmb\
\vardef\
}
	\label{fig:finalplot-sumet}
\end{figure}

\begin{figure}[h]
  \begin{center}
    \subfigure[]{
      \includegraphics[scale=0.36]{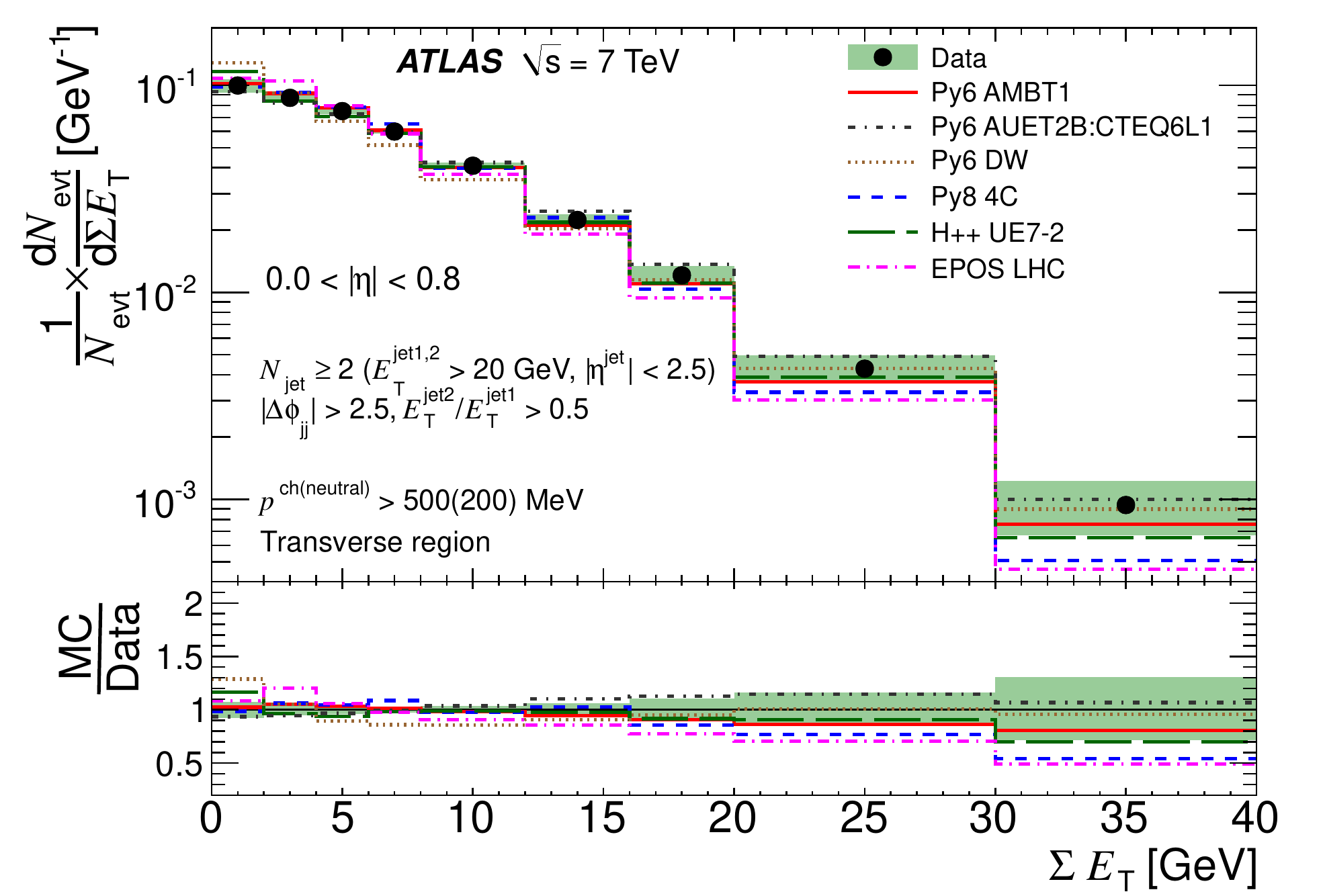}
    }
    \subfigure[]{
      \includegraphics[scale=0.36]{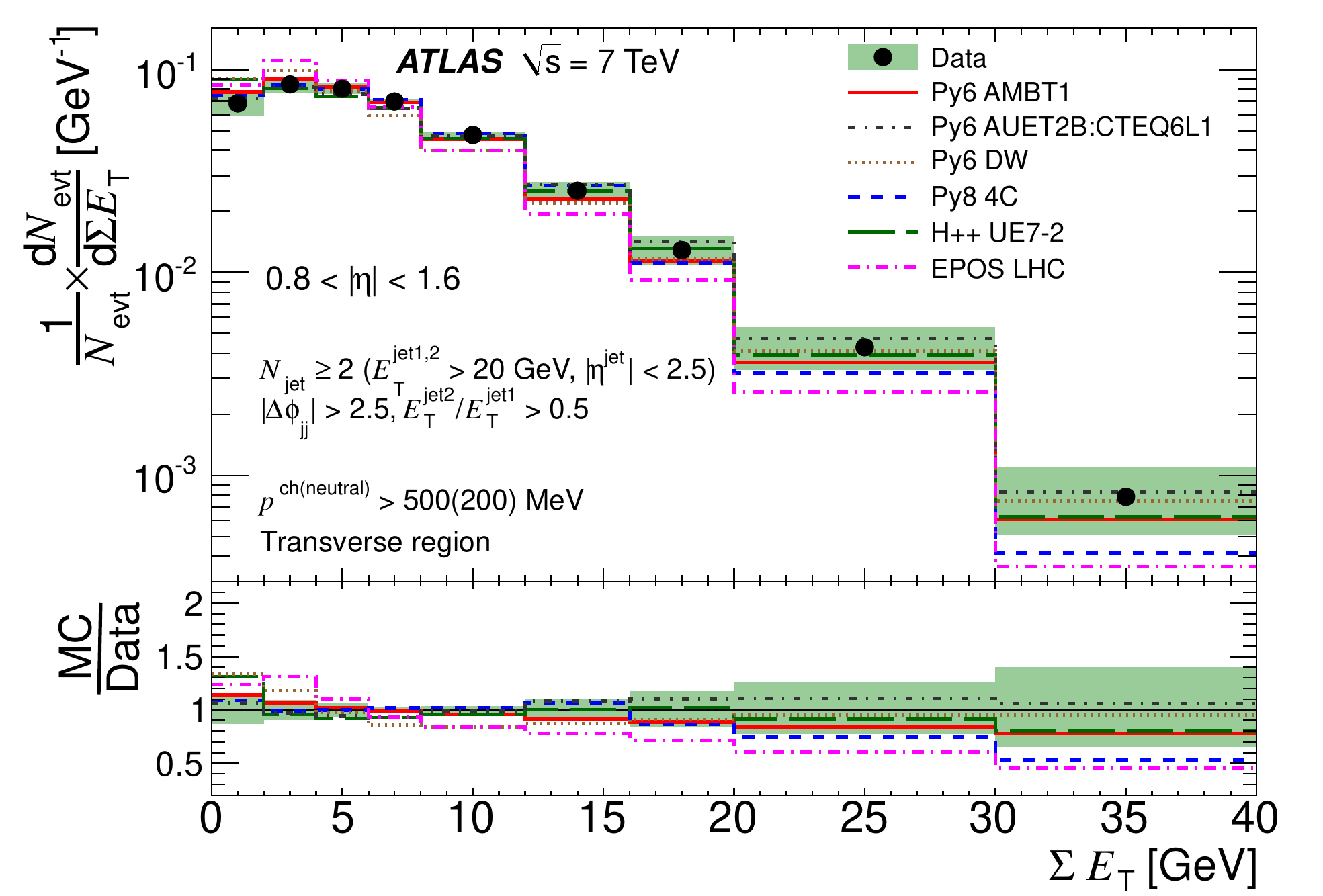}
    }
    \subfigure[]{
      \includegraphics[scale=0.36]{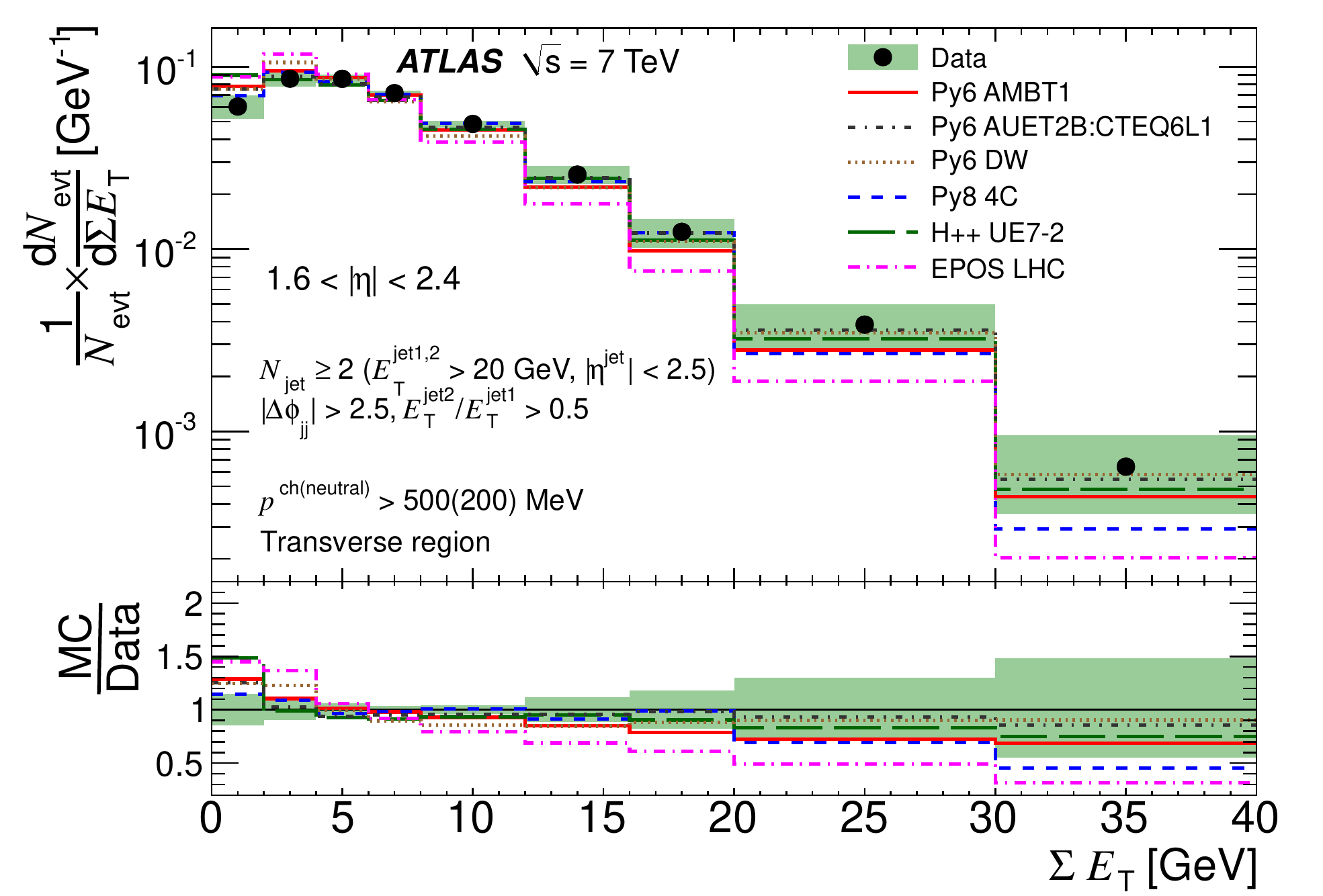}
    }
    \subfigure[]{
      \includegraphics[scale=0.36]{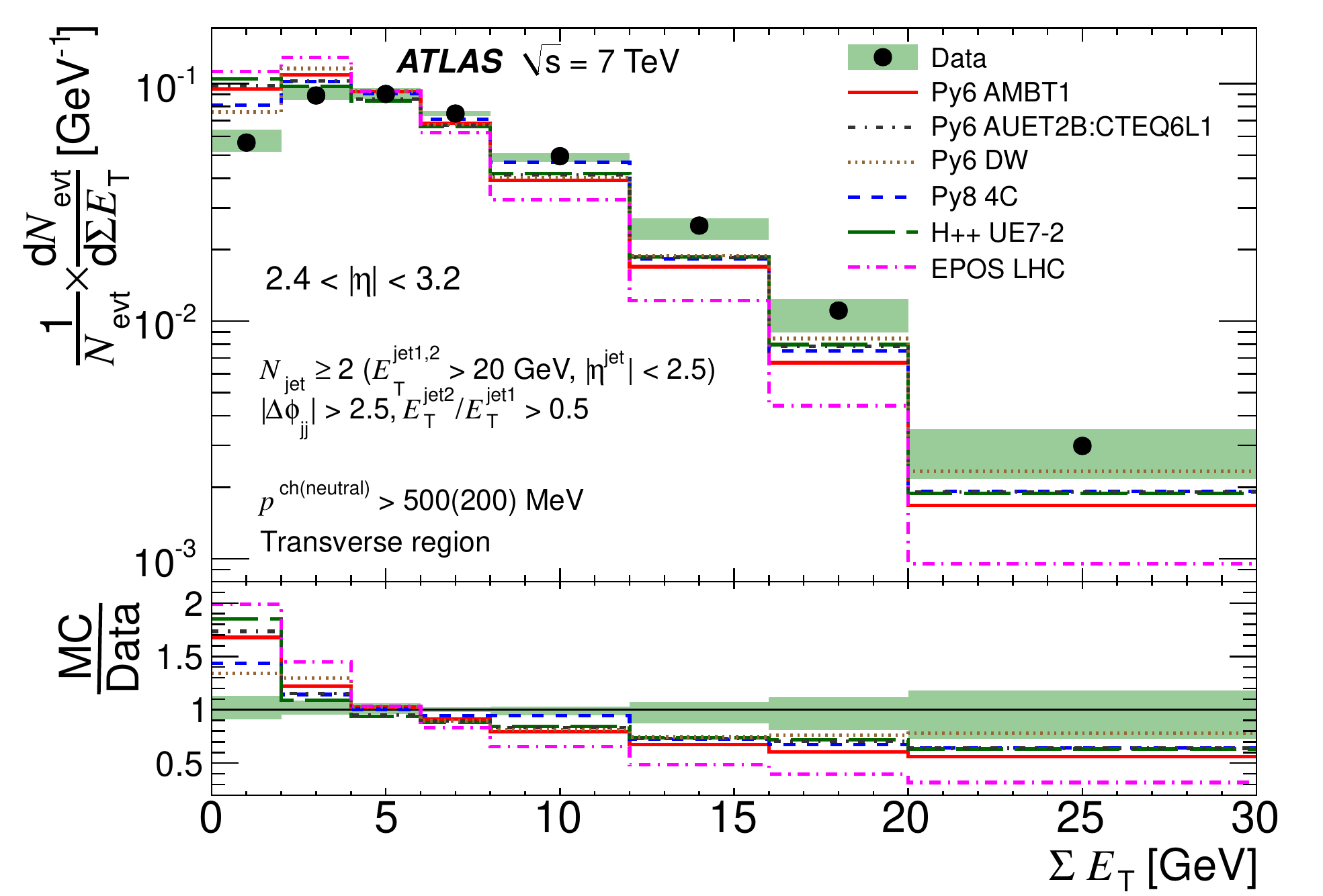}
    }
    \subfigure[]{
      \includegraphics[scale=0.36]{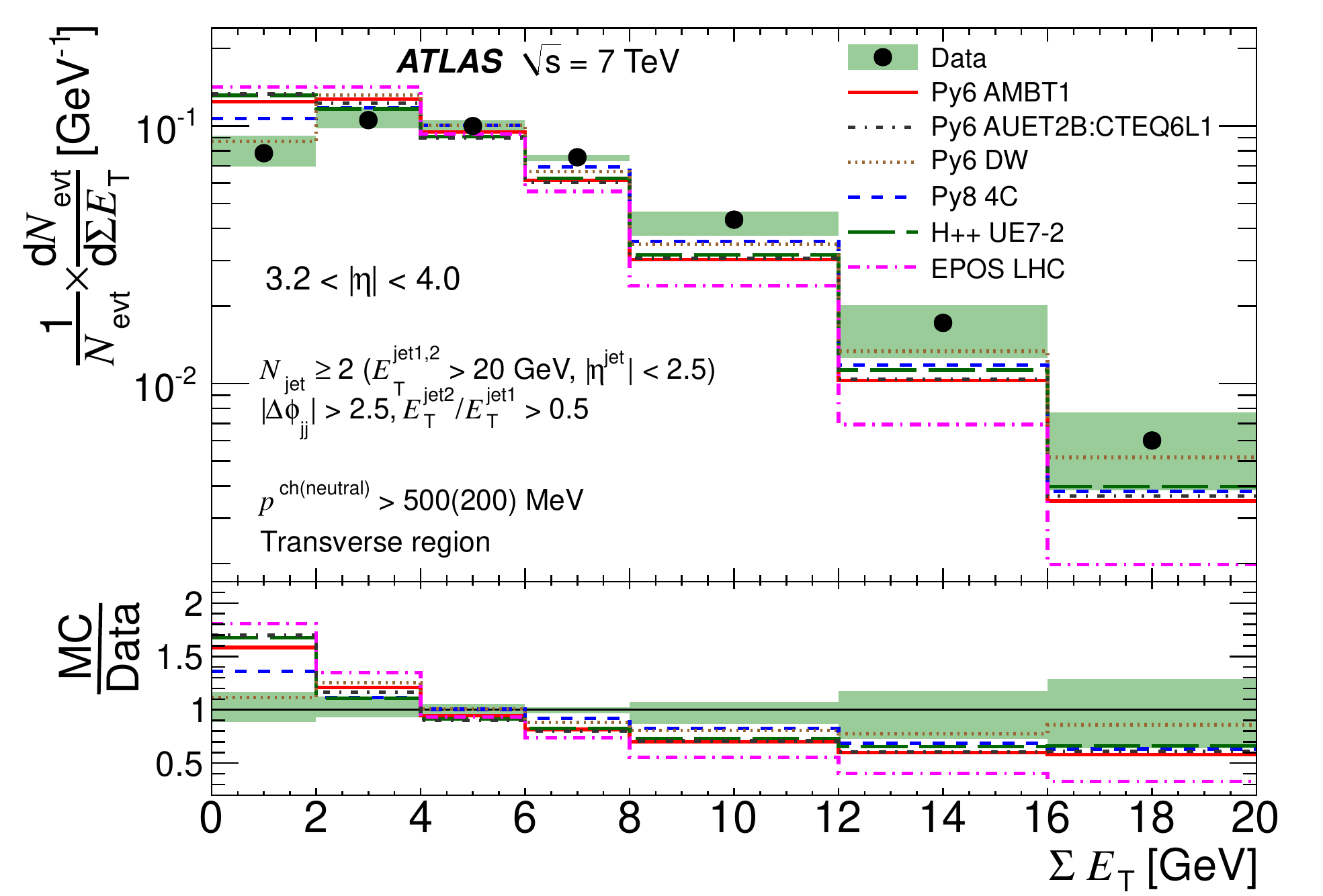}
    }
    \subfigure[]{
      \includegraphics[scale=0.36]{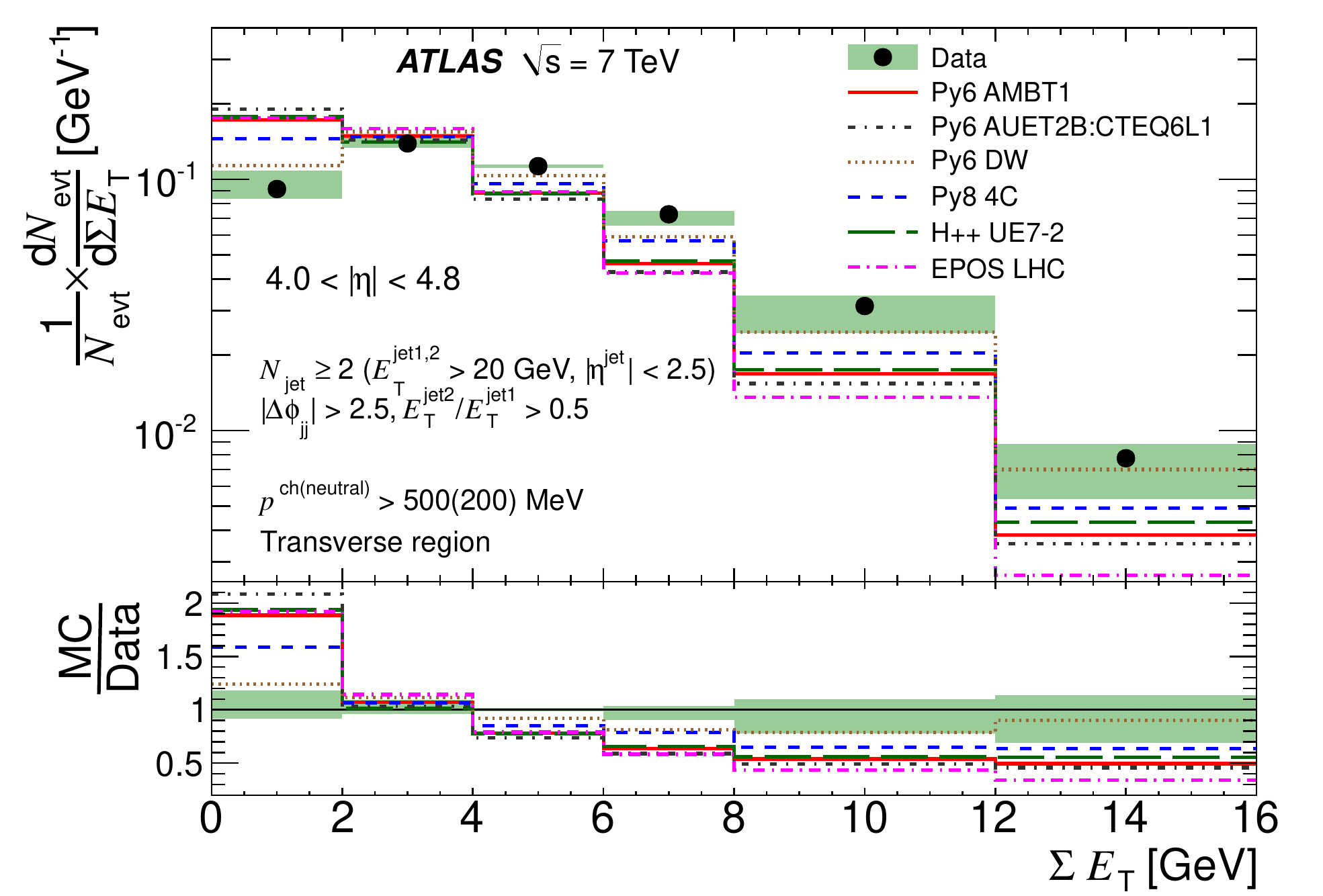}
    }
  \end{center}
  \caption{Unfolded $\sumet$ distributions compared to various MC models and tunes for the dijet selection in the transverse region in the following $|\eta|$ regions:
(a) \mbox{$0.0<|\eta|<0.8$},
(b) \mbox{$0.8<|\eta|<1.6$},
(c) \mbox{$1.6<|\eta|<2.4$}, 
(d) \mbox{$2.4<|\eta|<3.2$}, 
(e) \mbox{$3.2<|\eta|<4.0$} and 
(f) \mbox{$4.0<|\eta|<4.8$}. 
The filled band in each plot represents the total uncertainty on the unfolded data.
\vardefdj\
\vardef\
}
  \label{fig:ueplot-sumet}
\end{figure}

The  unfolded $\sumet$ distributions are shown in figures~\ref{fig:finalplot-sumet} and~\ref{fig:ueplot-sumet} for the minimum bias and dijet selections, respectively. The distribution peaks at higher values of $\sumet$ in the forward region due to the particle momentum cuts discussed above.
In the region \mbox{$|\eta|<3.2$} the distribution is broader than in the forward region, with more events populating the high $\sumet$ tails. 
There is therefore more event-by-event variation in the $\sumet$ in the central part of the detector. These features are reproduced by the MC predictions. 
\pythiasix\ \ambt\ provides the best description of the $\sumet$ shape in the central region for the minimum bias data. For the dijet data, most of the tunes do a reasonable job, although \pythiaeight\ \fourc\ and \eposlhc\ underestimate the high $\sumet$ tails.
As with the $\etflow$ distributions, the $\sumet$ in the forward region is underestimated for all but the dijet \pythiasix\ \DW\ prediction.

In summary, all of the MCs underestimate the amount of energy
in the forward region relative to the central region, in both the minimum bias data and the underlying event, with the exception of \pythiasix\ \DW\, which provides a reasonable description of the dijet data, although the prediction is approximately one standard deviation below the central values measured in the data in all $|\eta|$ bins.
\eposlhc\ provides the best overall description of the minimum bias data. \pythiasix\ \ambt\ provides the best description in the most central region \mbox{($|\eta\
|<1.6$)}, while at higher $|\eta|$ values \pythiaeight\ \fourc\ and \hpp\ \ueseven\ reflect the data more accurately. In the dijet analysis, all the MCs provide a reasonable description in the central region, apart from \eposlhc.

\subsection{Variation in diffractive contributions}

\begin{figure}[h]
	\begin{center}
          \includegraphics[scale=0.55]{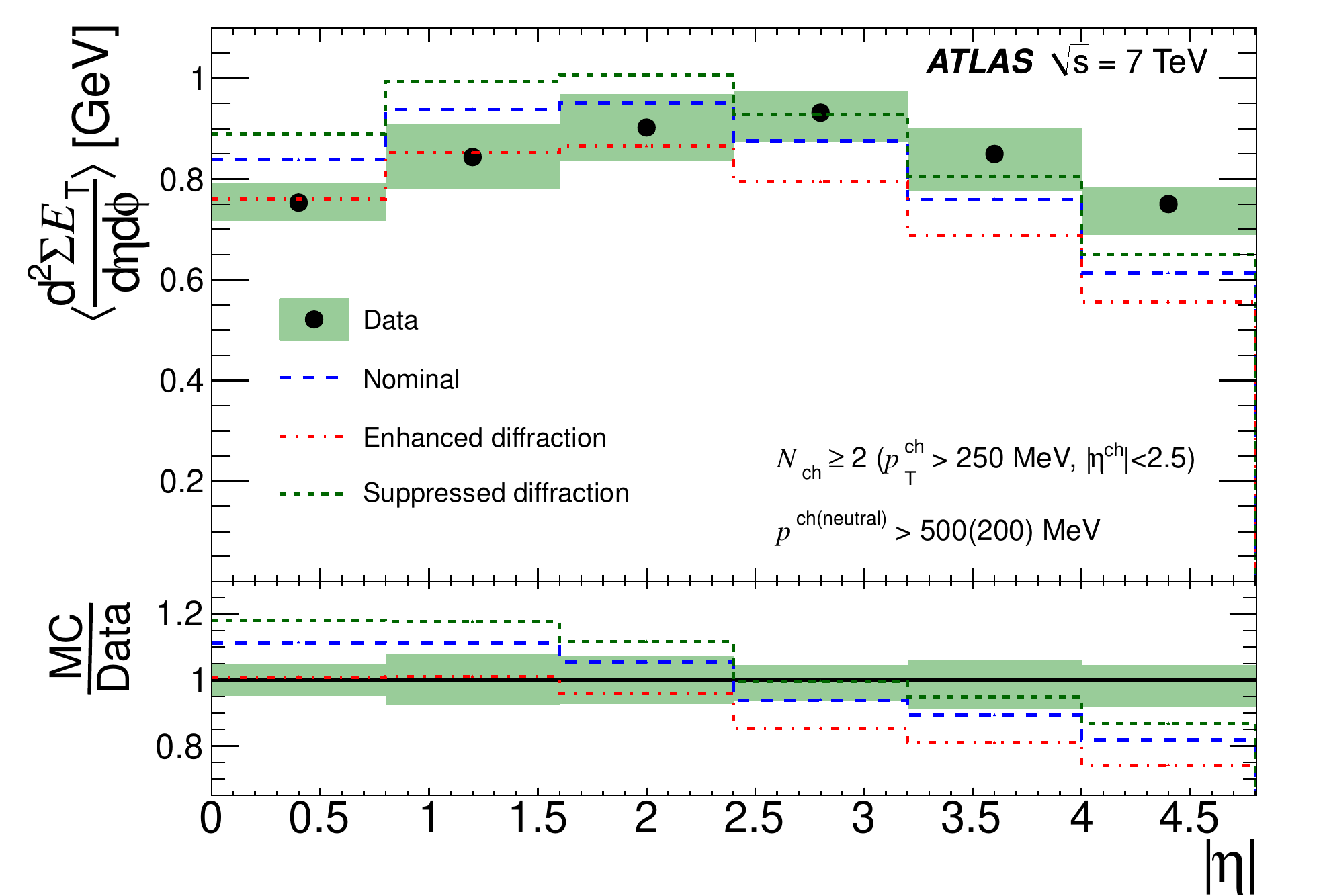}
	\end{center}
	\caption{Final unfolded $\etflow$ distribution for the minimum bias selection compared to
          \pythiaeight\ \fourc\ with the nominal diffractive
          cross-sections, as well as enhanced and suppressed diffractive cross-sections, as described in the text. The filled band represents the total uncertainty on the unfolded data. 
\vardefmb\
\vardef\
}
	\label{fig:finalplot-py8-sig}
\end{figure}

In order to investigate the sensitivity of the $\etflow$  to the
fraction of diffractive events, figure~\ref{fig:finalplot-py8-sig}
compares the unfolded $\etflow$ distribution in the minimum bias data to \pythiaeight\ \fourc\ with
the nominal diffractive cross-sections (50.9 mb, 12.4 mb and 8.1 mb for non-diffractive,
single-diffractive and double-diffractive processes, respectively) and to samples where the
diffractive cross-sections have been doubled or halved, with the non-diffractive cross-section held constant. This is achieved by combining the separate MC samples for the different processes with adjusted weights, rather than by changing the relevant parameters when generating the samples.
Diffractive processes tend to have less particle production than non-diffractive processes.
As expected, increasing the
diffractive contribution decreases the $\etflow$. However, the
shape of the $\etflow$ distribution is not significantly affected. 
\subsection{Variation in parton distribution functions}
\begin{figure}[h]
	\begin{center}
          \subfigure[]{       \includegraphics[scale=0.58]{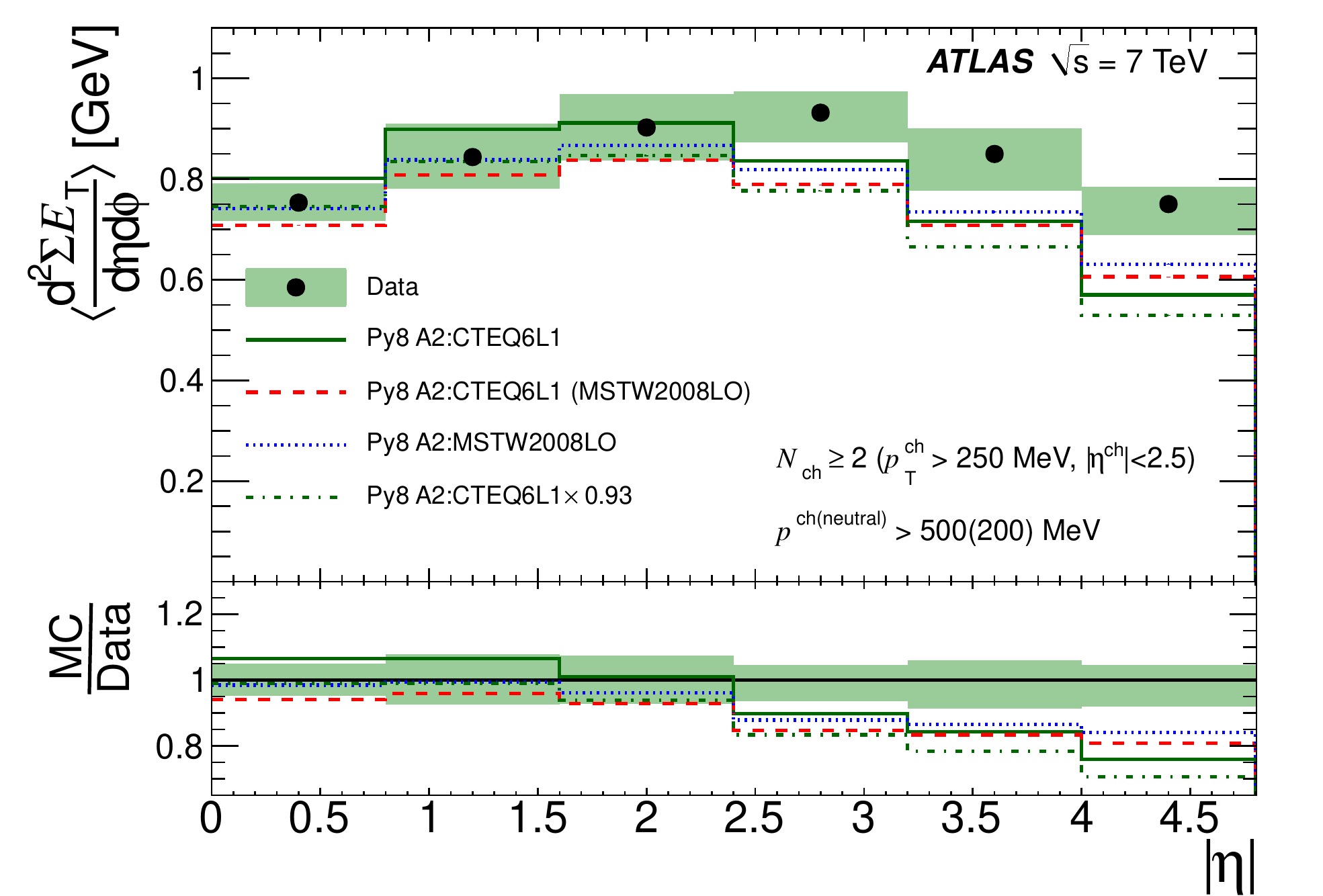}
	\label{fig:finalplot-py8-mb}
}
\subfigure[]{   
\includegraphics[scale=0.58]{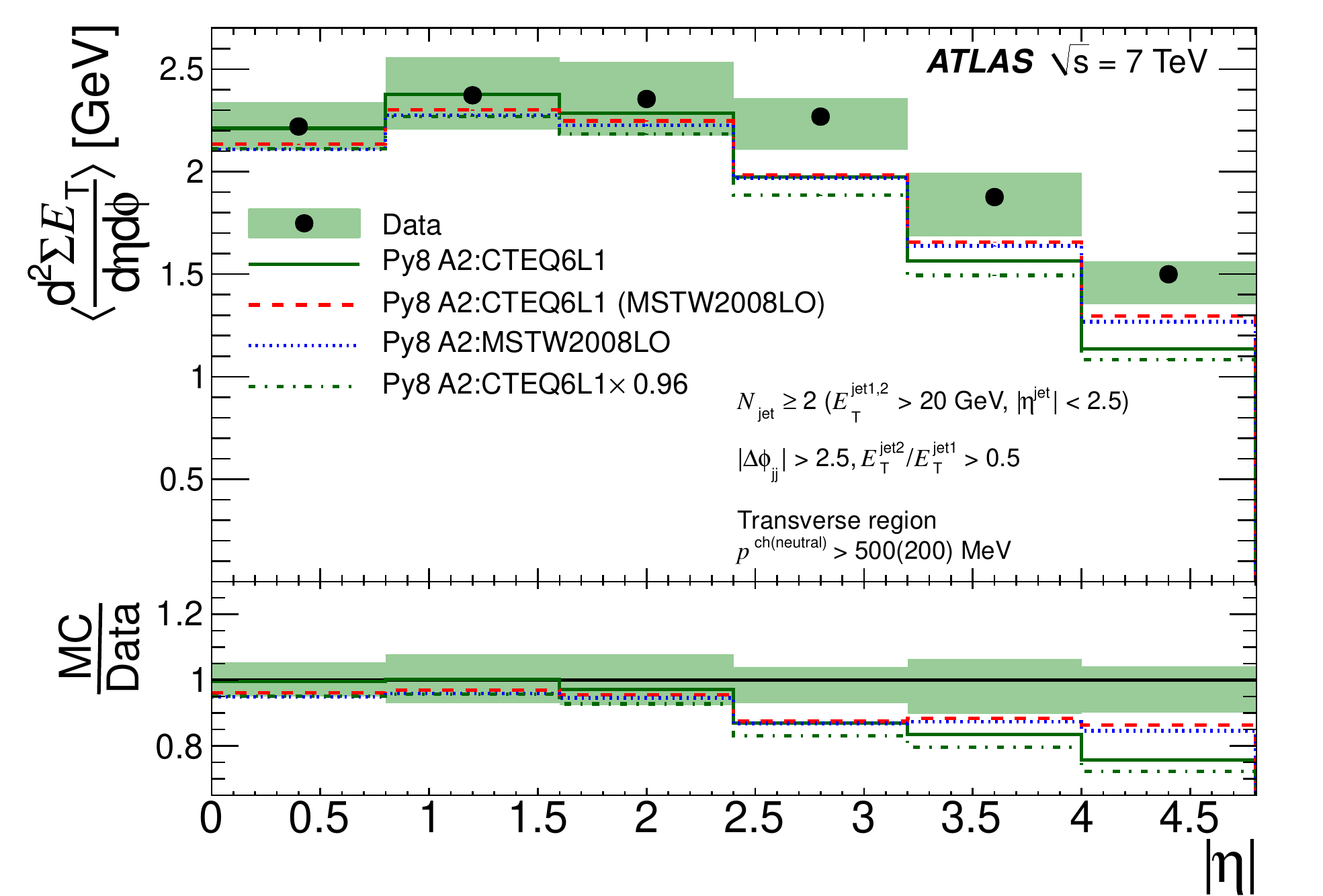}
	\label{fig:finalplot-py8-dj}
}
	\end{center}
	\caption{Final unfolded $\etflow$ distribution compared to \pythiaeight\ with variations of the PDFs used, as discussed in the text for (a) the minimum bias selection and (b) the dijet selection. The filled band represents the total uncertainty on the unfolded data. 
\vardefmb\
\vardefdj\
\vardef\
}
	\label{fig:finalplot-py8}
\end{figure}

The overall energy as well as its $|\eta|$ dependence are affected by the PDFs used as input to the MC model.
In order to investigate the dependence on the PDFs, comparisons are made between the data and the \pythiaeight\ \atwo\ family of tunes, which use different input PDFs~\cite{cite:atwo}, with the following variations:

\begin{enumerate}
\item Tune \atwoc.
\item The \atwoc\ tune parameters, but with the \mstwlo\ PDFs.
\item Tune \atwom.
\item Tune \atwoc\ where the $\etflow$ has been scaled by 0.93(0.96) for the minimum bias (dijet) selection so that it matches \atwom\ in the most central bin.
\end{enumerate}
These comparisons are shown in figure~\ref{fig:finalplot-py8}.
The first thing to note is that moving from the \cteqsix\ to the \mstwlo\ PDFs (and keeping all tune parameters the same) decreases the amount of energy in the central region, but increases it in the forward region, presumably due to the increase in both the high-$x$ and low-$x$ gluon PDF with respect to the mid-$x$ region, where $x$ is the proton momentum fraction carried by the gluon.
When the parameters are tuned to data in the central region, the energy increases for the minimum bias prediction.
If the $\etflow$ obtained using \atwoc\ is scaled down  to match \atwom\ in the most central bin,  it is clear that the latter provides a better description of the data in the forward region, with the underestimation in the most forward bin improving from about 30\%\ to 15\%. 

\section{Conclusions}
\label{sec:conclu}
Measurements of the $\etflow$ and the $\sumet$ distributions as functions of $|\eta|$ have been presented for two event classes: those requiring the presence of particles with a low transverse momentum (minimum bias) and those requiring particles with a significant transverse momentum (dijets),
using proton-proton collision data at \mbox{$\sqrt{s}=$7 TeV} recorded by the ATLAS detector. In the dijet selection the distributions are measured in the region transverse in $\phi$ to the hard scatter, in order to probe the particle production from the underlying event. The measurements are performed in the region \mbox{$|\eta|<4.8$} for charged particles with \mbox{$p>500$~MeV} and neutral particles with \mbox{$p>200$~MeV}, and are the first to utilize the entire acceptance of the ATLAS calorimeters to probe the overall properties of inclusive proton-proton collisions, as well as the underlying event. The distributions are compared to various MC models and tunes. In general all MC predictions are found to underestimate the amount of energy in the forward region relative to the central region by 20--30\%, with the exception of the \pythiasix\ \DW\ tune  and \eposlhc\ for the minimum bias data, although  \pythiasix\ \DW\ underpredicts the overall energy by 20--30\%. For the \pythiaeight\ \atwo\ tune series, this is improved if the \mstwlo\ PDFs are used instead of the \cteqsix\ PDFs.

\section{Acknowledgements}
We thank CERN for the very successful operation of the LHC, as well as the
support staff from our institutions without whom ATLAS could not be
operated efficiently.

We acknowledge the support of ANPCyT, Argentina; YerPhI, Armenia; ARC,
Australia; BMWF, Austria; ANAS, Azerbaijan; SSTC, Belarus; CNPq and FAPESP,
Brazil; NSERC, NRC and CFI, Canada; CERN; CONICYT, Chile; CAS, MOST and NSFC,
China; COLCIENCIAS, Colombia; MSMT CR, MPO CR and VSC CR, Czech Republic;
DNRF, DNSRC and Lundbeck Foundation, Denmark; EPLANET and ERC, European Union;
IN2P3-CNRS, CEA-DSM/IRFU, France; GNSF, Georgia; BMBF, DFG, HGF, MPG and AvH
Foundation, Germany; GSRT, Greece; ISF, MINERVA, GIF, DIP and Benoziyo Center,
Israel; INFN, Italy; MEXT and JSPS, Japan; CNRST, Morocco; FOM and NWO,
Netherlands; RCN, Norway; MNiSW, Poland; GRICES and FCT, Portugal; MERYS
(MECTS), Romania; MES of Russia and ROSATOM, Russian Federation; JINR; MSTD,
Serbia; MSSR, Slovakia; ARRS and MVZT, Slovenia; DST/NRF, South Africa;
MICINN, Spain; SRC and Wallenberg Foundation, Sweden; SER, SNSF and Cantons of
Bern and Geneva, Switzerland; NSC, Taiwan; TAEK, Turkey; STFC, the Royal
Society and Leverhulme Trust, United Kingdom; DOE and NSF, United States of
America.

The crucial computing support from all WLCG partners is acknowledged
gratefully, in particular from CERN and the ATLAS Tier-1 facilities at
TRIUMF (Canada), NDGF (Denmark, Norway, Sweden), CC-IN2P3 (France),
KIT/GridKA (Germany), INFN-CNAF (Italy), NL-T1 (Netherlands), PIC (Spain),
ASGC (Taiwan), RAL (UK) and BNL (USA) and in the Tier-2 facilities
worldwide.
\clearpage

\appendix
\section{Tabulated results and uncertainties}
\label{sec:tables}
The unfolded data are presented in tabular form in this appendix for the $\etflow$ and the six $\sumet$ distributions, for both the minimum bias and dijet selections.
Tables~\ref{tab:syst},~\ref{tab:syst-dijet} and~\ref{tab:syst-ratio} give the unfolded data and systematic uncertainties for the $\etflow$ for the minimum bias selection, the dijet selection, and the ratio between them, respectively. 
Tables~\ref{tab:syst1}--\ref{tab:syst6} give the unfolded data and systematic uncertainties for the $\sumet$ distributions for the minimum bias selection and tables~\ref{tab:systdijet1}--\ref{tab:systdijet6} give the corresponding information for the dijet analysis.
In each case, the breakdown of the systematic uncertainties by source is also given. Each systematic source is described in section~\ref{sec:systs}.
The uncorrelated calorimeter energy scale systematic is denoted as $E_{1}^{a,b,c,d,e,f}$ for each of the six $|\eta|$ regions, respectively. The correlated calorimeter energy scale systematic is denoted as $E_{2}$.
The two correlated material systematic sources are denoted as  $M_{1}$ and $M_{2}$ and the uncorrelated source is denoted as $M_{3}^{a,b,c,d,e,f}$ for the six $|\eta|$ regions. 
All the above sources are correlated between the minimum bias data and the dijet data and therefore have the same symbol.

The physics model systematic uncertainty on the minimum bias and dijet results, and on their ratio,  are denoted as, $P_{1}$, $P_{2}$ and $P$, respectively.
The jet energy scale systematic uncertainty is denoted as $J$. The physics model and jet energy scale systematic sources are uncorrelated between the minimum bias and dijet data. For the $\sumet$ distributions $J$ is negligible and therefore neglected in the region $|\eta| > 2.4$.

The correlations between bins of a given distribution are indicated by the sign of the uncertainty. For example, in table~\ref{tab:syst1} the uncertainty $E_{1}^{a}$ is $\pm$ in the first three bins and $\mp$ in the remaining bins. This means that the first three bins are correlated with each other and anti-correlated with the remaining bins (a downward shift in the $\sumet$ will shift the low $\sumet$ bins up and the high $\sumet$ bins down).
Since the individual sources within a given distribution are uncorrelated, the relationship between $\pm$ and $\mp$ between sources is not relevant to the calculation of the total error in a given bin.

The uncertainties are given to two significant figures or a precision of 0.01\%, whichever is smaller. In cases where the $+$ and $-$  uncertainty have a different precision the lowest precision is chosen for both. In cases where the uncertainty is not applicable, this is indicated with a dash.

\begin{table}
\begin{center}
	\begin{tabular}{lcllllllll}  
          \hline \hline
          $|\eta|$    & $\etflowlong$ &  Stat.   &  $E_{1}^{*}$  & $E_{2}  $   & $M_{1}$     & $M_{2}$   & $M_{3}^{*}$   & $P_1$    & Total   \\
                      & $[\rm GeV]$     &  $[\%]$   & $[\%]$          & $[\%]$        & $[\%]$       & $[\%]$      &  $[\%]$        & $[\%]$ & $[\%]$  \\
\hline
0.0  --  0.8  &  0.753   &   $\pm$0.19   &  $^{+3.2}_{-2.9}$   &  ---  &   $\pm$2.9  &   ---  &   $\pm$0.51  &   $\pm$2.6   &  $^{+5.1}_{-4.9}$ \vspace{1mm} \\
0.8  --  1.6  &  0.844   &   $\pm$0.17   &  $^{+5.4}_{-4.9}$   &  ---  &   $\pm$3.2  &   $\pm$0.49  &   $\pm$1.2  &   $\pm$4.6   &  $^{+7.9}_{-7.5}$ \vspace{1mm} \\
1.6  --  2.4  &  0.902   &   $\pm$0.16   &  $^{+4.0}_{-3.8}$   &  ---  &   ---  &   $\pm$0.89  &   $\pm$5.0  &   $\pm$3.4   &  $^{+7.4}_{-7.2}$ \vspace{1mm} \\
2.4  --  3.2  &  0.932   &   $\pm$0.16   &  $^{+2.4}_{-5.0}$   &  ---  &   ---  &  ---   &   $\pm$3.0  &   $\pm$2.5   &  $^{+4.6}_{-6.4}$ \vspace{1mm} \\
3.2  --  4.0  &  0.850   &   $\pm$0.15   &  $^{+4.3}_{-4.4}$   &  $-6.2$  &  ---   &  ---  &   $\pm$2.7  &   $\pm$3.2   &  $^{+6.0}_{-8.7}$ \vspace{1mm} \\
4.0  --  4.8  &  0.750   &   $\pm$0.14   &  $^{+2.7}_{-2.7}$   &  $-6.8$  &  ---  &  ---  &   $\pm$0.8  &   $\pm$3.6   &  $^{+4.6}_{-8.2}$ \vspace{1mm} \\
          \hline \hline
	\end{tabular}
	\caption{Measured $\etflow$ and systematic uncertainty breakdown for the minimum bias data. The systematic uncertainties marked with a $*$ are uncorrelated between $|\eta|$ bins.}
        \label{tab:syst}
\end{center}
\end{table}

\begin{table}
\begin{center}
	\begin{tabular}{lclllllllll} 
          \hline \hline
          $|\eta|$    & $\etflowlong$ &  Stat.   &  $E_{1}^{*}$  & $E_{2}  $   & $M_{1}$     & $M_{2}$   & $M_{3}^{*}$   & $P_2$  & $J$   & Total   \\
                      & $[\rm GeV]$     &  $[\%]$    & $[\%]$         & $[\%]$        & $[\%]$       & $[\%]$      &  $[\%]$        & $[\%]$  & $[\%]$   & $[\%]$ \\
\hline
0.0  --  0.8  &  2.22   &   $\pm$0.61   &  $^{+4.3}_{-4.2}$   &  ---  &   $\pm$1.3  &   ---  &   $\pm$0.23  &   $\pm$2.2  &   $^{+1.6}_{-1.3}$   &  $^{+5.3}_{-5.1}$ \vspace{1mm} \\
0.8  --  1.6  &  2.37   &   $\pm$0.54   &  $^{+7.2}_{-6.4}$   &  ---  &   $\pm$2.5  &   $\pm$0.38  &   $\pm$0.96  &   $\pm$0.12  &   $^{+1.3}_{-1.3}$   &  $^{+7.8}_{-7.1}$ \vspace{1mm} \\
1.6  --  2.4  &  2.35   &   $\pm$0.52   &  $^{+5.3}_{-5.0}$   &  ---  &   ---  &   $\pm$0.97  &   $\pm$5.5  &   $\pm$0.41  &   $^{+0.98}_{-0.92}$   &  $^{+7.8}_{-7.6}$ \vspace{1mm} \\
2.4  --  3.2  &  2.27   &   $\pm$0.50   &  $^{+3.8}_{-7.0}$   &  ---  &   ---  &   ---  &   $\pm$0.64  &   $\pm$0.55  &   $^{+0.80}_{-0.37}$   &  $^{+4.0}_{-7.1}$ \vspace{1mm} \\
3.2  --  4.0  &  1.88   &   $\pm$0.51   &  $^{+6.1}_{-5.8}$   &  $-8.2$  &  ---  &   ---  &   $\pm$1.1  &   $\pm$1.3  &   $^{+0.46}_{-0.17}$   &  $^{+6}_{-10}$ \vspace{1mm} \\
4.0  --  4.8  &  1.50   &   $\pm$0.47   &  $^{+3.8}_{-3.6}$   &  $-9.0$  &  ---  &   ---  &   $\pm$0.6  &   $\pm$1.6  &   $^{+0.13}_{-0.03}$   &  $^{+4.2}_{-9.8}$ \vspace{1mm} \\
          \hline \hline
	\end{tabular}
	\caption{Measured $\etflow$ and systematic uncertainty breakdown for the dijet data. The systematic uncertainties marked with a $*$ are uncorrelated between $|\eta|$ bins.}
        \label{tab:syst-dijet}
\end{center}
\end{table}

\begin{table}
\begin{center}
	\begin{tabular}{lclllllllll} 
          \hline \hline
          $|\eta|$    & $\frac{\etflowlong(UE)}{\etflowlong(MB)}$ &  Stat.   &  $E_{1}^{*}$  & $E_{2}  $   & $M_{1}$     & $M_{2}$   & $M_{3}^{*}$   & $P$  & $J$   & Total   \\
                      &       &  $[\%]$    & $[\%]$         & $[\%]$        & $[\%]$       & $[\%]$      &  $[\%]$        & $[\%]$  & $[\%]$   & $[\%]$ \\
\hline

0.0  --  0.8  &  2.95   &   $\pm$0.64   &  $^{+1.1}_{-1.3}$   &  ---      &  $^{+1.5}_{-1.6}$   &  ---   &  $^{+0.27}_{-0.28}$  &   $\pm$3.4  &  $^{+1.6}_{-1.3}$   &  $^{+4.3}_{-4.2}$ \vspace{1mm} \\
0.8  --  1.6  &  2.81   &   $\pm$0.57   &  $^{+1.7}_{-1.6}$   &  ---      &  $^{+0.64}_{-0.69}$   &  $^{+0.10}_{-0.11}$   &  $^{+0.25}_{-0.26}$  &   $\pm$4.6  &  $^{+1.3}_{-1.3}$   &  $^{+5.1}_{-5.1}$ \vspace{1mm} \\
1.6  --  2.4  &  2.61   &   $\pm$0.55   &  $^{+1.2}_{-1.2}$   &  ---      &  ---   &  $^{+0.08}_{-0.08}$   &  $^{+0.43}_{-0.47}$  &   $\pm$3.5  &  $^{+0.98}_{-0.92}$   &  $^{+3.9}_{-3.9}$ \vspace{1mm} \\
2.4  --  3.2  &  2.43   &   $\pm$0.52   &  $^{+1.4}_{-2.1}$   &  ---      &  ---  &  ---   &  $^{+2.3}_{-2.5}$  &   $\pm$2.6  &  $^{+0.80}_{-0.37}$   &  $^{+3.8}_{-4.2}$ \vspace{1mm} \\
3.2  --  4.0  &  2.21   &   $\pm$0.53   &  $^{+1.7}_{-1.6}$   &  $-2.2$   &  ---   &  ---   &  $^{+1.5}_{-1.6}$  &   $\pm$3.4  &  $^{+0.46}_{-0.17}$   &  $^{+4.1}_{-4.6}$ \vspace{1mm} \\
4.0  --  4.8  &  2.00   &   $\pm$0.49   &  $^{+1.1}_{-1.0}$   &  $-2.4$  &  ---   &  ---   &  $^{+0.19}_{-0.20}$  &   $\pm$3.9  &  $^{+0.13}_{-0.03}$   &  $^{+4.1}_{-4.7}$ \vspace{1mm} \\

     \hline \hline
	\end{tabular}
	\caption{Ratio of measured $\etflow$ for the dijet data to that for the the minimum bias data, and systematic uncertainty breakdown. The systematic uncertainties marked with a $*$ are uncorrelated between $|\eta|$ bins.}
        \label{tab:syst-ratio}
\end{center}
\end{table}

\begin{table}
\begin{center}
	\begin{tabular}{lrllllllll} 
          \hline \hline
 $\sumet$   & $\sumetlong$  & Stat.   &  $E_{1}^{a}$  & $E_{2}  $   & $ M_{1}$     & $ M_{2}$   & $ M_{3}^{a}$   & $P_1$    & Total   \\
  $[\rm GeV]$     &   $[\rm GeV^{-1}]$   &  $[\%]$   & $[\%]$     & $[\%]$  & $[\%]$  & $[\%]$ &  $[\%]$ & $[\%]$ & $[\%]$  \\

\hline 

0  --  2  &  0.161   &   0.35 &  $^{+2.0}_{-2.1}$   &  ---  &   $\mp$ 2.1  &   --- &   $\mp$ 0.38  &   $\pm$0.07   &   $^{+3.0}_{-3.1}$  \vspace{1mm} \\
2  --  4  &  0.0835   &   0.32 &  $^{+1.3}_{-1.4}$   &  ---  &   $\mp$ 1.2  &   --- &   $\mp$ 0.21  &   $\mp$4.5   &   $^{+4.8}_{-4.8}$  \vspace{1mm} \\
4  --  6  &  0.0533   &   0.40 &  $^{+0.24}_{-0.39}$   &  ---  &   $\mp$ 0.90  &   --- &   $\mp$ 0.16  &   $\mp$3.8   &   $^{+3.9}_{-3.9}$  \vspace{1mm} \\
6  --  8  &  0.039   &   0.46 &  $^{-0.20}_{+0.00}$   &  ---  &   $\pm$ 0.23  &   --- &   $\pm$ 0.04  &   $\mp$0.44   &   $^{+0.68}_{-0.71}$  \vspace{1mm} \\
8  --  12  &  0.0271   &   0.49 &  $^{-0.57}_{+0.56}$   &  ---  &   $\pm$ 1.4  &   --- &   $\pm$ 0.24  &   $\pm$2.2   &   $^{+2.7}_{-2.7}$  \vspace{1mm} \\
12  --  16  &  0.0177   &   0.57 &  $^{-1.3}_{+1.5}$   &  ---  &   $\pm$ 1.8  &   --- &   $\pm$ 0.31  &   $\pm$3.5   &   $^{+4.3}_{-4.2}$  \vspace{1mm} \\
16  --  20  &  0.0121   &   0.67 &  $^{-2.5}_{+2.5}$   &  ---  &   $\pm$ 2.4  &   --- &   $\pm$ 0.43  &   $\pm$3.6   &   $^{+5.1}_{-5.1}$  \vspace{1mm} \\
20  --  30  &  0.00619   &   0.75 &  $^{-4.6}_{+4.8}$   &  ---  &   $\pm$ 4.1  &   --- &   $\pm$ 0.73  &   $\pm$4.2   &   $^{+7.7}_{-7.5}$  \vspace{1mm} \\
30  --  40  &  0.00226   &   1.2 &  $^{-7.5}_{+8.4}$   &  ---  &   $\pm$ 7.9  &   --- &   $\pm$ 1.4  &   $\pm$5.5   &   $^{+13}_{-12}$  \vspace{1mm} \\
40  --  50  &  0.000855   &   1.9 &  $^{-10}_{+12}$   &  ---  &   $\pm$ 8.7  &   --- &   $\pm$ 1.5  &   $\pm$8.1   &   $^{+17}_{-16}$  \vspace{1mm} \\
50  --  60  &  0.000321   &   2.5 &  $^{-13}_{+15}$   &  ---  &   $\pm$ 13  &   --- &   $\pm$ 2.3  &   $\pm$10   &   $^{+22}_{-21}$  \vspace{1mm} \\

          \hline \hline
	\end{tabular}
	\caption{\tablecaption\ minimum bias data in the region \mbox{$0.0<|\eta|<0.8$}.}
        \label{tab:syst1}
\end{center}
\end{table}

\begin{table}
\begin{center}
	\begin{tabular}{lrllllllll} 
          \hline \hline

 $\sumet$   & $\sumetlong$  & Stat.   &  $ E_{1}^{b}$  & $ E_{2}  $   & $ M_{1}$     & $ M_{2}$   & $ M_{3}^{b}$   & $P_1$    & Total   \\
  $[\rm GeV]$     &   $[\rm GeV^{-1}]$              &  $[\%]$   & $[\%]$     & $[\%]$  & $[\%]$  & $[\%]$ &  $[\%]$ & $[\%]$ & $[\%]$  \\

\hline

0  --  2  &  0.123   &   0.38 &  $^{+4.7}_{-4.8}$   &  ---  &   $\mp$ 3.0  &   $\mp$ 0.46  &   $\mp$ 1.1  &   $\pm$0.64   &   $^{+5.7}_{-5.8}$  \vspace{1mm} \\
2  --  4  &  0.092   &   0.30 &  $^{+2.5}_{-2.9}$   &  ---  &   $\mp$ 1.6  &   $\mp$ 0.25  &   $\mp$ 0.62  &   $\mp$5.3   &   $^{+6.1}_{-6.2}$  \vspace{1mm} \\
4  --  6  &  0.0588   &   0.35 &  $^{+0.55}_{-0.85}$   &  ---  &   $\mp$ 0.65  &   $\mp$ 0.1  &   $\mp$ 0.25  &   $\mp$9   &   $^{+9.0}_{-9.1}$  \vspace{1mm} \\
6  --  8  &  0.0425   &   0.42 &  $^{-0.29}_{+0.21}$   &  ---  &   $\pm$ 0.07  &   $\pm$ 0.01  &   $\pm$ 0.03  &   $\mp$3.8   &   $^{+3.8}_{-3.8}$  \vspace{1mm} \\
8  --  12  &  0.0296   &   0.44 &  $^{-1.0}_{+1.1}$   &  ---  &   $\pm$ 0.76  &   $\pm$ 0.12  &   $\pm$ 0.29  &   $\pm$1.3   &   $^{+1.9}_{-1.9}$  \vspace{1mm} \\
12  --  16  &  0.0198   &   0.51 &  $^{-2.2}_{+2.0}$   &  ---  &   $\pm$ 1.9  &   $\pm$ 0.29  &   $\pm$ 0.73  &   $\pm$5.7   &   $^{+6.4}_{-6.4}$  \vspace{1mm} \\
16  --  20  &  0.0137   &   0.58 &  $^{-3.9}_{+3.8}$   &  ---  &   $\pm$ 2.1  &   $\pm$ 0.32  &   $\pm$ 0.80  &   $\pm$8.3   &   $^{+9.4}_{-9.4}$  \vspace{1mm} \\
20  --  30  &  0.00726   &   0.67 &  $^{-7.3}_{+7.6}$   &  ---  &   $\pm$ 4.7  &   $\pm$ 0.72  &   $\pm$ 1.8  &   $\pm$10   &   $^{+14}_{-13}$  \vspace{1mm} \\
30  --  40  &  0.00268   &   1.1 &  $^{-13}_{+14}$   &  ---  &   $\pm$ 7.3  &   $\pm$ 1.1  &   $\pm$ 2.8  &   $\pm$9   &   $^{+19}_{-17}$  \vspace{1mm} \\
40  --  50  &  0.000951   &   1.7 &  $^{-18}_{+21}$   &  ---  &   $\pm$ 13  &   $\pm$ 1.9  &   $\pm$ 4.9  &   $\pm$10   &   $^{+27}_{-25}$  \vspace{1mm} \\
50  --  60  &  0.000339   &   2.4 &  $^{-22}_{+30}$   &  ---  &   $\pm$ 17  &   $\pm$ 2.6  &   $\pm$ 6.5  &   $\pm$15   &   $^{+38}_{-32}$  \vspace{1mm} \\
          \hline \hline
	\end{tabular}
	\caption{\tablecaption\ minimum bias data in the region \mbox{$0.8<|\eta|<1.6$}.}
        \label{tab:syst2}
\end{center}
\end{table}

\begin{table}
\begin{center}
	\begin{tabular}{lrllllllll} 
          \hline \hline

 $\sumet$   & $\sumetlong$  & Stat.   &  $ E_{1}^{c}$  & $ E_{2}  $   & $ M_{1}$     & $ M_{2}$   & $ M_{3}^{c}$   & $P_1$    & Total   \\
  $[\rm GeV]$     &    $[\rm GeV^{-1}]$             &  $[\%]$   & $[\%]$     & $[\%]$  & $[\%]$  & $[\%]$ &  $[\%]$ & $[\%]$ & $[\%]$  \\

\hline
0  --  2  &  0.0980   &   0.41 &  $^{+4.4}_{-4.3}$   &  ---  &   --- &   $\mp$ 1.2  &   $\mp$ 6.8  &   $\pm$1.1   &   $^{+8.2}_{-8.2}$  \vspace{1mm} \\
2  --  4  &  0.0931   &   0.32 &  $^{+2.5}_{-2.6}$   &  ---  &   --- &   $\mp$ 0.48  &   $\mp$ 2.7  &   $\mp$2.0   &   $^{+4.2}_{-4.3}$  \vspace{1mm} \\
4  --  6  &  0.0639   &   0.36 &  $^{+0.57}_{-0.76}$   &  ---  &   --- &   $\mp$ 0.18  &   $\mp$ 1.0  &   $\mp$6.2   &   $^{+6.3}_{-6.4}$  \vspace{1mm} \\
6  --  8  &  0.0460   &   0.45 &  $^{-0.15}_{+0.05}$   &  ---  &   --- &   $\pm$ 0.27  &   $\pm$ 1.6  &   $\mp$4.8   &   $^{+5.1}_{-5.1}$  \vspace{1mm} \\
8  --  12  &  0.0323   &   0.46 &  $^{-0.61}_{+0.55}$   &  ---  &   --- &   $\pm$ 0.26  &   $\pm$ 1.5  &   $\mp$0.75   &   $^{+1.8}_{-1.8}$  \vspace{1mm} \\
12  --  16  &  0.0216   &   0.54 &  $^{-1.6}_{+1.5}$   &  ---  &   --- &   $\pm$ 0.33  &   $\pm$ 1.9  &   $\pm$2.1   &   $^{+3.2}_{-3.3}$  \vspace{1mm} \\
16  --  20  &  0.0149   &   0.62 &  $^{-2.9}_{+2.7}$   &  ---  &   --- &   $\pm$ 0.59  &   $\pm$ 3.4  &   $\pm$3.5   &   $^{+5.6}_{-5.7}$  \vspace{1mm} \\
20  --  30  &  0.00792   &   0.68 &  $^{-5.6}_{+5.6}$   &  ---  &   --- &   $\pm$ 1.2  &   $\pm$ 7.1  &   $\pm$6.8   &   $^{+11}_{-11}$  \vspace{1mm} \\
30  --  40  &  0.00290   &   1.1 &  $^{-11}_{+11}$   &  ---  &   --- &   $\pm$ 2.6  &   $\pm$ 15  &   $\pm$12   &   $^{+22}_{-22}$  \vspace{1mm} \\
40  --  50  &  0.000977   &   1.8 &  $^{-15}_{+17}$   &  ---  &   --- &   $\pm$ 3.2  &   $\pm$ 18  &   $\pm$14   &   $^{+29}_{-28}$  \vspace{1mm} \\
50  --  60  &  0.000312   &   2.5 &  $^{-19}_{+24}$   &  ---  &   --- &   $\pm$ 4.3  &   $\pm$ 25  &   $\pm$14   &   $^{+37}_{-34}$  \vspace{1mm} \\

\hline \hline
	\end{tabular}
	\caption{\tablecaption\ minimum bias data in the region \mbox{$1.6<|\eta|<2.4$}.}
        \label{tab:syst3}
\end{center}
\end{table}

\begin{table}
\begin{center}
	\begin{tabular}{lrllllllll} 
          \hline \hline

 $\sumet$   & $\sumetlong$  & Stat.   &  $ E_{1}^{d}$  & $ E_{2}  $   & $ M_{1}$     & $ M_{2}$   & $ M_{3}^{d}$   & $P_1$    & Total   \\
  $[\rm GeV]$     &  $[\rm GeV^{-1}]$               &  $[\%]$   & $[\%]$     & $[\%]$  & $[\%]$  & $[\%]$ &  $[\%]$ & $[\%]$ & $[\%]$  \\

\hline 

0  --  2  &  0.0840   &   0.48 &  $^{+6.9}_{-3.1}$   &  ---  &   --- &   --- &   $\mp$ 5.3  &   $\pm$0.72   &   $^{+8.8}_{-6.2}$  \vspace{1mm} \\
2  --  4  &  0.0920   &   0.33 &  $^{+3.1}_{-1.6}$   &  ---  &   --- &   --- &   $\mp$ 1.7  &   $\mp$1.5   &   $^{+3.9}_{-2.8}$  \vspace{1mm} \\
4  --  6  &  0.0660   &   0.46 &  $^{+0.65}_{-0.41}$   &  ---  &   --- &   --- &   $\pm$ 0.46  &   $\mp$5.3   &   $^{+5.3}_{-5.3}$  \vspace{1mm} \\
6  --  8  &  0.0485   &   0.47 &  $^{-0.16}_{+0.07}$   &  ---  &   --- &   --- &   $\pm$ 0.47  &   $\mp$3.5   &   $^{+3.5}_{-3.5}$  \vspace{1mm} \\
8  --  12  &  0.0343   &   0.50 &  $^{-0.46}_{+0.25}$   &  ---  &   --- &   --- &   $\pm$ 0.33  &   $\mp$0.57   &   $^{+0.86}_{-0.95}$  \vspace{1mm} \\
12  --  16  &  0.0232   &   0.57 &  $^{-1.5}_{+0.6}$   &  ---  &   --- &   --- &   $\pm$ 1.1  &   $\pm$3.1   &   $^{+3.4}_{-3.7}$  \vspace{1mm} \\
16  --  20  &  0.0160   &   0.61 &  $^{-3.7}_{+1.6}$   &  ---  &   --- &   --- &   $\pm$ 2.7  &   $\pm$3.3   &   $^{+4.6}_{-5.7}$  \vspace{1mm} \\
20  --  30  &  0.00829   &   0.72 &  $^{-7.8}_{+3.6}$   &  ---  &   --- &   --- &   $\pm$ 4.6  &   $\pm$5.3   &   $^{+8}_{-11}$  \vspace{1mm} \\
30  --  40  &  0.00290   &   1.2 &  $^{-14}_{+7}$   &  ---  &   --- &   --- &   $\pm$ 7.6  &   $\pm$6.9   &   $^{+12}_{-17}$  \vspace{1mm} \\
40  --  50  &  0.000908   &   1.6 &  $^{-22}_{+12}$   &  ---  &   --- &   --- &   $\pm$ 12  &   $\pm$11   &   $^{+20}_{-27}$  \vspace{1mm} \\
50  --  60  &  0.000281   &   3.2 &  $^{-27}_{+14}$   &  ---  &   --- &   --- &   $\pm$ 15  &   $\pm$7.3   &   $^{+22}_{-32}$  \vspace{1mm} \\

          \hline \hline
	\end{tabular}
	\caption{\tablecaption\ minimum bias data in the region \mbox{$2.4<|\eta|<3.2$}.}
        \label{tab:syst4}
\end{center}
\end{table}

\begin{table}
\begin{center}
	\begin{tabular}{lrllllllll} 
          \hline \hline

 $\sumet$   & $\sumetlong$  & Stat.   &  $ E_{1}^{e}$  & $ E_{2}  $   & $ M_{1}$     & $ M_{2}$   & $ M_{3}^{e}$   & $P_1$    & Total   \\
  $[\rm GeV]$     &    $[\rm GeV^{-1}]$             &  $[\%]$   & $[\%]$     & $[\%]$  & $[\%]$  & $[\%]$ &  $[\%]$ & $[\%]$ & $[\%]$  \\

\hline 

0  --  2  &  0.0837   &   0.46 &  $^{+5.2}_{-4.3}$   &  $7.3$  &   --- &   --- &   $\mp$ 4.0  &   $\mp$1.5   &   $^{+9.9}_{-6.1}$  \vspace{1mm} \\
2  --  4  &  0.0931   &   0.33 &  $^{+3.5}_{-3.4}$   &  $4.9$  &   --- &   --- &   $\mp$ 2.1  &   $\pm$0.64   &   $^{+6.4}_{-4.0}$  \vspace{1mm} \\
4  --  6  &  0.0720   &   0.38 &  $^{+1.1}_{-1.3}$   &  $1.6$  &   --- &   --- &   $\mp$ 0.18  &   $\mp$4.1   &   $^{+4.5}_{-4.3}$  \vspace{1mm} \\
6  --  8  &  0.0530   &   0.44 &  $^{+0.23}_{-0.30}$   &  $0.33$  &   --- &   --- &   $\mp$ 0.57  &   $\mp$4.5   &   $^{+4.6}_{-4.6}$  \vspace{1mm} \\
8  --  12  &  0.0367   &   0.47 &  $^{-0.69}_{+0.36}$   &  $-0.98$  &   --- &   --- &   $\pm$ 1.6  &   $\mp$2.2   &   $^{+2.8}_{-3.0}$  \vspace{1mm} \\
12  --  16  &  0.0237   &   0.54 &  $^{-2.4}_{+1.9}$   &  $-3.3$  &   --- &   --- &   $\pm$ 1.1  &   $\pm$2.3   &   $^{+3.2}_{-4.9}$  \vspace{1mm} \\
16  --  20  &  0.0154   &   0.66 &  $^{-4.6}_{+3.8}$   &  $-6.4$  &   --- &   --- &   $\pm$ 2.6  &   $\pm$5.3   &   $^{+7.0}_{-9.9}$  \vspace{1mm} \\
20  --  30  &  0.00704   &   0.77 &  $^{-9.3}_{+8.6}$   &  $-13$  &   --- &   --- &   $\pm$ 5.7  &   $\pm$9.5   &   $^{+14}_{-20}$  \vspace{1mm} \\
30  --  40  &  0.00183   &   1.4 &  $^{-18}_{+19}$   &  $-25$  &   --- &   --- &   $\pm$ 7.5  &   $\pm$13   &   $^{+24}_{-34}$  \vspace{1mm} \\
40  --  50  &  0.000385   &   2.4 &  $^{-24}_{+31}$   &  $-33$  &   --- &   --- &   $\pm$ 20  &   $\pm$13   &   $^{+39}_{-48}$  \vspace{1mm} \\

          \hline \hline
	\end{tabular}
	\caption{\tablecaption\ minimum bias data in the region \mbox{$3.2<|\eta|<4.0$}.}
        \label{tab:syst5}
\end{center}
\end{table}

\begin{table}
\begin{center}
	\begin{tabular}{lrllllllll} 
          \hline \hline

 $\sumet$   & $\sumetlong$  & Stat.   &  $ E_{1}^{f}$  & $ E_{2}  $   & $ M_{1}$     & $ M_{2}$   & $ M_{3}^{f}$   & $P_1$    & Total   \\
  $[\rm GeV]$     &   $[\rm GeV^{-1}]$              &  $[\%]$   & $[\%]$     & $[\%]$  & $[\%]$  & $[\%]$ &  $[\%]$ & $[\%]$ & $[\%]$  \\

\hline 

0  --  2  &  0.0818   &   0.54 &  $^{+3.6}_{-3.2}$   &  $9.0$  &   --- &   --- &   $\mp$ 1.7  &   $\mp$0.37   &   $^{+9.9}_{-3.7}$  \vspace{1mm} \\
2  --  4  &  0.102   &   0.42 &  $^{+2.0}_{-2.0}$   &  $5.1$  &   --- &   --- &   $\mp$ 0.58  &   $\mp$1.1   &   $^{+5.7}_{-2.4}$  \vspace{1mm} \\
4  --  6  &  0.0777   &   0.44 &  $^{+0.60}_{-0.77}$   &  $1.5$  &   --- &   --- &   $\pm$ 0.54  &   $\mp$5.3   &   $^{+5.6}_{-5.4}$  \vspace{1mm} \\
6  --  8  &  0.0577   &   0.48 &  $^{+0.06}_{-0.04}$   &  $0.16$  &   --- &   --- &   $\mp$ 0.14  &   $\mp$4.5   &   $^{+4.5}_{-4.5}$  \vspace{1mm} \\
8  --  12  &  0.0397   &   0.49 &  $^{-0.60}_{+0.43}$   &  $-1.5$  &   --- &   --- &   $\pm$ 0.50  &   $\mp$0.26   &   $^{+0.9}_{-1.8}$  \vspace{1mm} \\
12  --  16  &  0.0239   &   0.58 &  $^{-2.4}_{+2.0}$   &  $-6.0$  &   --- &   --- &   $\pm$ 0.33  &   $\pm$3.6   &   $^{+4.1}_{-7.5}$  \vspace{1mm} \\
16  --  20  &  0.0135   &   0.75 &  $^{-4.8}_{+4.5}$   &  $-12$  &   --- &   --- &   $\pm$ 1.6  &   $\pm$9.1   &   $^{+10}_{-16}$  \vspace{1mm} \\
20  --  30  &  0.00464   &   0.96 &  $^{-8.7}_{+8.9}$   &  $-22$  &   --- &   --- &   $\pm$ 2.5  &   $\pm$14   &   $^{+17}_{-28}$  \vspace{1mm} \\
30  --  40  &  0.000642   &   2.3 &  $^{-15}_{+19}$   &  $-39$  &   --- &   --- &   $\pm$ 3.4  &   $\pm$24   &   $^{+31}_{-48}$  \vspace{1mm} \\

          \hline \hline
	\end{tabular}
	\caption{\tablecaption\ minimum bias data in the region \mbox{$4.0<|\eta|<4.8$}.}
        \label{tab:syst6}
\end{center}
\end{table}


\begin{table}
\begin{center}
	\begin{tabular}{lrlllllllll} 
          \hline \hline

 $\sumet$   & $\sumetlong$  & Stat.   &  $E_{1}^{a}$  & $E_{2}  $   & $ M_{1}$     & $ M_{2}$   & $ M_{3}^{a}$   & $P_{2}$    & $J$ & Total   \\
  $[\rm GeV]$     &   $[\rm GeV^{-1}]$              &  $[\%]$   & $[\%]$     & $[\%]$  & $[\%]$  & $[\%]$ &  $[\%]$ & $[\%]$ & $[\%]$  & $[\%]$ \\

\hline

0  --  2  &  0.0998   &   2.1 &  $^{+4.3}_{-4.4}$   &  ---  &   $\mp$ 2.1  &   --- &   $\mp$ 0.37  &   $\mp$5.1 &  $^{-1.2}_{+1.2}$   &   $^{+7.4}_{-7.4}$  \vspace{1mm} \\
2  --  4  &  0.0870   &   1.5 &  $^{+3.4}_{-3.7}$   &  ---  &   $\mp$ 1.4  &   --- &   $\mp$ 0.24  &   $\pm$2.6 &  $^{-0.61}_{+0.35}$   &   $^{+4.8}_{-5.0}$  \vspace{1mm} \\
4  --  6  &  0.0751   &   1.4 &  $^{+2.0}_{-1.8}$   &  ---  &   $\mp$ 0.67  &   --- &   $\mp$ 0.12  &   $\pm$2.7 &  $^{-0.38}_{+0.08}$   &   $^{+3.7}_{-3.6}$  \vspace{1mm} \\
6  --  8  &  0.0598   &   1.6 &  $^{+0.46}_{-0.59}$   &  ---  &   $\pm$ 0.05  &   --- &   $\pm$ 0.01  &   $\mp$1.8 &  $^{-0.34}_{+0.14}$   &   $^{+2.5}_{-2.5}$  \vspace{1mm} \\
8  --  12  &  0.0409   &   1.7 &  $^{-1.9}_{+1.9}$   &  ---  &   $\pm$ 1.1  &   --- &   $\pm$ 0.20  &   $\mp$2.0 &  $^{-0.23}_{+0.25}$   &   $^{+3.4}_{-3.4}$  \vspace{1mm} \\
12  --  16  &  0.0224   &   2.2 &  $^{-4.9}_{+5.3}$   &  ---  &   $\pm$ 2.5  &   --- &   $\pm$ 0.45  &   $\mp$0.15 &  $^{+0.14}_{+0.21}$   &   $^{+6.3}_{-6.0}$  \vspace{1mm} \\
16  --  20  &  0.0121   &   3.0 &  $^{-7.7}_{+7.0}$   &  ---  &   $\pm$ 4.0  &   --- &   $\pm$ 0.70  &   $\pm$6.0 &  $^{+1.0}_{-0.8}$   &   $^{+11}_{-11}$  \vspace{1mm} \\
20  --  30  &  0.00428   &   4.3 &  $^{-11}_{+13}$   &  ---  &   $\pm$ 6.5  &   --- &   $\pm$ 1.1  &   $\pm$3.9 &  $^{+5.4}_{-4.5}$   &   $^{+16}_{-15}$  \vspace{1mm} \\
30  --  40  &  0.000939   &   8.7 &  $^{-16}_{+18}$   &  ---  &   $\pm$ 10  &   --- &   $\pm$ 1.8  &   $\pm$16 &  $^{+13}_{-11}$   &   $^{+31}_{-28}$  \vspace{1mm} \\

          \hline \hline
	\end{tabular}
	\caption{\tablecaption\ dijet data in the region \mbox{$0.0<|\eta|<0.8$}.}
        \label{tab:systdijet1}
\end{center}
\end{table}

\begin{table}
\begin{center}
	\begin{tabular}{lrlllllllll} 
          \hline \hline

 $\sumet$   & $\sumetlong$  & Stat.   &  $E_{1}^{b}$  & $E_{2}  $   & $ M_{1}$     & $ M_{2}$   & $ M_{3}^{b}$   & $P_{2}$    & $J$ & Total   \\
  $[\rm GeV]$     &     $[\rm GeV^{-1}]$            &  $[\%]$   & $[\%]$     & $[\%]$  & $[\%]$  & $[\%]$ &  $[\%]$ & $[\%]$ & $[\%]$   & $[\%]$\\

\hline

0  --  2  &  0.0680   &   2.2 &  $^{+9.7}_{-9.6}$   &  ---  &   $\mp$ 4.0  &   $\mp$ 0.62  &   $\mp$ 1.5  &   $\mp$7.7 &  $^{-0.79}_{+0.89}$   &   $^{+13}_{-13}$  \vspace{1mm} \\
2  --  4  &  0.0844   &   1.5 &  $^{+7.1}_{-7.4}$   &  ---  &   $\mp$ 2.8  &   $\mp$ 0.44  &   $\mp$ 1.1  &   $\pm$4.8 &  $^{-0.80}_{+0.81}$   &   $^{+9.2}_{-9.4}$  \vspace{1mm} \\
4  --  6  &  0.0804   &   1.4 &  $^{+4.1}_{-4.4}$   &  ---  &   $\mp$ 1.7  &   $\mp$ 0.25  &   $\mp$ 0.64  &   $\pm$3.7 &  $^{-0.83}_{+0.91}$   &   $^{+6.0}_{-6.2}$  \vspace{1mm} \\
6  --  8  &  0.0695   &   1.4 &  $^{+0.9}_{-1.6}$   &  ---  &   $\mp$ 0.48  &   $\mp$ 0.07  &   $\mp$ 0.19  &   $\pm$2.0 &  $^{-0.62}_{+0.69}$   &   $^{+2.7}_{-3.0}$  \vspace{1mm} \\
8  --  12  &  0.0476   &   1.6 &  $^{-3.4}_{+2.7}$   &  ---  &   $\pm$ 1.3  &   $\pm$ 0.20  &   $\pm$ 0.49  &   $\mp$0.86 &  $^{-0.01}_{+0.09}$   &   $^{+3.5}_{-4.1}$  \vspace{1mm} \\
12  --  16  &  0.0252   &   2.0 &  $^{-8.9}_{+8.9}$   &  ---  &   $\pm$ 3.6  &   $\pm$ 0.56  &   $\pm$ 1.4  &   $\mp$4.2 &  $^{+0.69}_{-1.1}$   &   $^{+11}_{-11}$  \vspace{1mm} \\
16  --  20  &  0.0129   &   2.8 &  $^{-13}_{+15}$   &  ---  &   $\pm$ 6.0  &   $\pm$ 0.92  &   $\pm$ 2.3  &   $\mp$4.8 &  $^{+1.3}_{-2.3}$   &   $^{+17}_{-16}$  \vspace{1mm} \\
20  --  30  &  0.00429   &   4.2 &  $^{-19}_{+23}$   &  ---  &   $\pm$ 10  &   $\pm$ 1.6  &   $\pm$ 3.9  &   $\pm$3.5 &  $^{+4.2}_{-4.2}$   &   $^{+26}_{-23}$  \vspace{1mm} \\
30  --  40  &  0.000785   &   9.5 &  $^{-26}_{+32}$   &  ---  &   $\pm$ 16  &   $\pm$ 2.5  &   $\pm$ 6.1  &   $\pm$6.9 &  $^{+13}_{-10}$   &   $^{+40}_{-35}$  \vspace{1mm} \\

          \hline \hline
	\end{tabular}
	\caption{\tablecaption\ dijet data in the region \mbox{$0.8<|\eta|<1.6$}.}
        \label{tab:systdijet2}
\end{center}
\end{table}

\begin{table}
\begin{center}
	\begin{tabular}{lrlllllllll} 
          \hline \hline

 $\sumet$   & $\sumetlong$  & Stat.   &  $E_{1}^{c}$  & $E_{2}  $   & $ M_{1}$     & $ M_{2}$   & $ M_{3}^{c}$   & $P_{2}$    & $J$ & Total   \\
  $[\rm GeV]$     &      $[\rm GeV^{-1}]$           &  $[\%]$   & $[\%]$     & $[\%]$  & $[\%]$  & $[\%]$ &  $[\%]$ & $[\%]$ & $[\%]$ & $[\%]$  \\

\hline

0  --  2  &  0.0604   &   2.4 &  $^{+7.6}_{-7.5}$   &  ---  &   --- &   $\mp$ 1.6  &   $\mp$ 9.2  &   $\mp$8.2 &  $^{+0.01}_{+0.52}$   &   $^{+15}_{-15}$  \vspace{1mm} \\
2  --  4  &  0.0859   &   1.6 &  $^{+6.1}_{-6.0}$   &  ---  &   --- &   $\mp$ 1.2  &   $\mp$ 6.6  &   $\pm$2.6 &  $^{-0.43}_{+0.51}$   &   $^{+9.6}_{-9.5}$  \vspace{1mm} \\
4  --  6  &  0.0857   &   1.5 &  $^{+3.4}_{-3.3}$   &  ---  &   --- &   $\mp$ 0.70  &   $\mp$ 4.0  &   $\pm$2.1 &  $^{-0.47}_{+0.27}$   &   $^{+5.8}_{-5.8}$  \vspace{1mm} \\
6  --  8  &  0.0715   &   1.6 &  $^{+0.8}_{-1.3}$   &  ---  &   --- &   $\mp$ 0.24  &   $\mp$ 1.3  &   $\pm$2.5 &  $^{-0.40}_{+0.22}$   &   $^{+3.3}_{-3.5}$  \vspace{1mm} \\
8  --  12  &  0.0485   &   1.7 &  $^{-2.7}_{+2.2}$   &  ---  &   --- &   $\pm$ 0.46  &   $\pm$ 2.6  &   $\pm$0.51 &  $^{-0.63}_{+0.33}$   &   $^{+3.9}_{-4.2}$  \vspace{1mm} \\
12  --  16  &  0.0256   &   2.3 &  $^{-7.3}_{+7.4}$   &  ---  &   --- &   $\pm$ 1.4  &   $\pm$ 7.9  &   $\mp$3 &  $^{-0.30}_{+0.10}$   &   $^{+12}_{-11}$  \vspace{1mm} \\
16  --  20  &  0.0124   &   3.3 &  $^{-12}_{+12}$   &  ---  &   --- &   $\pm$ 2.3  &   $\pm$ 13  &   $\mp$1.2 &  $^{+0.75}_{-0.51}$   &   $^{+18}_{-18}$  \vspace{1mm} \\
20  --  30  &  0.00385   &   4.8 &  $^{-17}_{+18}$   &  ---  &   --- &   $\pm$ 4.0  &   $\pm$ 22  &   $\pm$1.3 &  $^{+4.7}_{-3.9}$   &   $^{+30}_{-29}$  \vspace{1mm} \\
30  --  40  &  0.000641   &   11 &  $^{-20}_{+25}$   &  ---  &   --- &   $\pm$ 6.3  &   $\pm$ 36  &   $\mp$2.0 &  $^{+17}_{-14}$   &   $^{+48}_{-45}$  \vspace{1mm} \\

          \hline \hline
	\end{tabular}
	\caption{\tablecaption\ dijet data in the region \mbox{$1.6<|\eta|<2.4$}.}
        \label{tab:systdijet3}
\end{center}
\end{table}

\begin{table}
\begin{center}
	\begin{tabular}{lrllllllll} 
          \hline \hline

 $\sumet$   & $\sumetlong$  & Stat.   &  $E_{1}^{d}$  & $E_{2}  $   & $ M_{1}$     & $ M_{2}$   & $ M_{3}^{d}$   & $P_{2}$    & Total   \\
  $[\rm GeV]$     &      $[\rm GeV^{-1}]$           &  $[\%]$   & $[\%]$     & $[\%]$  & $[\%]$  & $[\%]$ &  $[\%]$ & $[\%]$ & $[\%]$   \\

\hline

0  --  2  &  0.0565   &   2.9 &  $^{+11}_{-5}$   &  ---  &   --- &   --- &   $\mp$ 2.6  &   $\mp$6.0   &   $^{+13}_{-9}$  \vspace{1mm} \\
2  --  4  &  0.0892   &   1.7 &  $^{+8.5}_{-3.8}$   &  ---  &   --- &   --- &   $\mp$ 1.8  &   $\pm$0.53   &   $^{+8.9}_{-4.5}$  \vspace{1mm} \\
4  --  6  &  0.0905   &   1.7 &  $^{+5.1}_{-2.5}$   &  ---  &   --- &   --- &   $\mp$ 0.9  &   $\pm$2.7   &   $^{+6.1}_{-4.2}$  \vspace{1mm} \\
6  --  8  &  0.0748   &   1.7 &  $^{+1.1}_{-1.0}$   &  ---  &   --- &   --- &   $\mp$ 0.03  &   $\pm$1.3   &   $^{+2.4}_{-2.3}$  \vspace{1mm} \\
8  --  12  &  0.0495   &   1.8 &  $^{-4.6}_{+1.1}$   &  ---  &   --- &   --- &   $\pm$ 1.3  &   $\mp$1.2   &   $^{+2.8}_{-5.3}$  \vspace{1mm} \\
12  --  16  &  0.0253   &   2.6 &  $^{-12}_{+6}$   &  ---  &   --- &   --- &   $\pm$ 3.0  &   $\pm$1.5   &   $^{+7}_{-13}$  \vspace{1mm} \\
16  --  20  &  0.0111   &   3.5 &  $^{-18}_{+10}$   &  ---  &   --- &   --- &   $\pm$ 4.8  &   $\mp$2.1   &   $^{+12}_{-19}$  \vspace{1mm} \\
20  --  30  &  0.00298   &   5.8 &  $^{-25}_{+15}$   &  ---  &   --- &   --- &   $\pm$ 7.8  &   $\pm$1.2   &   $^{+18}_{-27}$  \vspace{1mm} \\

          \hline \hline
	\end{tabular}
	\caption{\tablecaption\ dijet data in the region \mbox{$2.4<|\eta|<3.2$}.}
        \label{tab:systdijet4}
\end{center}
\end{table}

\begin{table}
\begin{center}
	\begin{tabular}{lrllllllll} 
          \hline \hline

 $\sumet$   & $\sumetlong$  & Stat.   &  $E_{1}^{e}$  & $E_{2}  $   & $ M_{1}$     & $ M_{2}$   & $ M_{3}^{e}$   & $P_{2}$    &  Total   \\
  $[\rm GeV]$     &    $[\rm GeV^{-1}]$             &  $[\%]$   & $[\%]$     & $[\%]$  & $[\%]$  & $[\%]$ &  $[\%]$ & $[\%]$ & $[\%]$   \\

\hline 

0  --  2  &  0.0784   &   2.5 &  $^{+8.3}_{-7.3}$   &  $12$  &   --- &   --- &   $\mp$ 5.1  &   $\mp$6.6   &   $^{+17}_{-11}$  \vspace{1mm} \\
2  --  4  &  0.105   &   1.7 &  $^{+6.7}_{-6.2}$   &  $9.5$  &   --- &   --- &   $\mp$ 2.9  &   $\pm$1.7   &   $^{+12}_{-7.3}$  \vspace{1mm} \\
4  --  6  &  0.0999   &   1.6 &  $^{+2.8}_{-3.1}$   &  $3.9$  &   --- &   --- &   $\mp$ 0.8  &   $\pm$2.0   &   $^{+5.5}_{-4.1}$  \vspace{1mm} \\
6  --  8  &  0.0756   &   1.7 &  $^{-1.4}_{+0.3}$   &  $-2.0$  &   --- &   --- &   $\pm$ 1.3  &   $\pm$0.47   &   $^{+2.2}_{-3.3}$  \vspace{1mm} \\
8  --  12  &  0.0433   &   2.0 &  $^{-7.1}_{+5.6}$   &  $-10$  &   --- &   --- &   $\pm$ 4.5  &   $\pm$0.46   &   $^{+8}_{-13}$  \vspace{1mm} \\
12  --  16  &  0.0172   &   2.9 &  $^{-15}_{+15}$   &  $-21$  &   --- &   --- &   $\pm$ 8.8  &   $\pm$0.25   &   $^{+17}_{-27}$  \vspace{1mm} \\
16  --  20  &  0.00601   &   4.7 &  $^{-19}_{+25}$   &  $-27$  &   --- &   --- &   $\pm$ 13  &   $\mp$0.76   &   $^{+29}_{-36}$  \vspace{1mm} \\

          \hline \hline
	\end{tabular}
	\caption{\tablecaption\ dijet data in the region \mbox{$3.2<|\eta|<4.0$}.}
        \label{tab:systdijet4}
\end{center}
\end{table}

\begin{table}
\begin{center}
	\begin{tabular}{lrllllllll} 
          \hline \hline

 $\sumet$   & $\sumetlong$  & Stat.   &  $E_{1}^{f}$  & $E_{2}  $   & $ M_{1}$     & $ M_{2}$   & $ M_{3}^{f}$   & $P_{2}$    & Total   \\
  $[\rm GeV]$     &    $[\rm GeV^{-1}]$             &  $[\%]$   & $[\%]$     & $[\%]$  & $[\%]$  & $[\%]$ &  $[\%]$ & $[\%]$ & $[\%]$   \\

\hline 

0  --  2  &  0.0915   &   2.5 &  $^{+6.2}_{-5.4}$   &  $16$  &   --- &   --- &   $\mp$ 0.63  &   $\mp$6.2   &   $^{+18}_{-9}$  \vspace{1mm} \\
2  --  4  &  0.139   &   1.6 &  $^{+3.4}_{-3.5}$   &  $8.5$  &   --- &   --- &   $\mp$ 0.12  &   $\pm$1.3   &   $^{+9.3}_{-4.1}$  \vspace{1mm} \\
4  --  6  &  0.113   &   1.6 &  $^{-0.13}_{-0.51}$   &  $-0.33$  &   --- &   --- &   $\pm$ 0.38  &   $\pm$0.38   &   $^{+1.7}_{-1.7}$  \vspace{1mm} \\
6  --  8  &  0.0726   &   1.8 &  $^{-3.6}_{+2.7}$   &  $-9.0$  &   --- &   --- &   $\pm$ 0.88  &   $\mp$0.18   &   $^{+3.4}_{-9.8}$  \vspace{1mm} \\
8  --  12  &  0.0313   &   2.5 &  $^{-7.9}_{+8.4}$   &  $-20$  &   --- &   --- &   $\pm$ 1.6  &   $\pm$4.7   &   $^{+10}_{-22}$  \vspace{1mm} \\
12  --  16  &  0.00775   &   4.4 &  $^{-11}_{+13}$   &  $-28$  &   --- &   --- &   $\pm$ 2.6  &   $\pm$2.2   &   $^{+14}_{-31}$  \vspace{1mm} \\

          \hline \hline
	\end{tabular}
	\caption{\tablecaption\ dijet data in the region \mbox{$4.0<|\eta|<4.8$}.}
        \label{tab:systdijet6}
\end{center}
\end{table}

\clearpage
\providecommand{\href}[2]{#2}\begingroup\raggedright\endgroup

\onecolumn
\clearpage
\begin{flushleft}
{\Large The ATLAS Collaboration}

\bigskip

G.~Aad$^{\rm 47}$,
T.~Abajyan$^{\rm 20}$,
B.~Abbott$^{\rm 110}$,
J.~Abdallah$^{\rm 11}$,
S.~Abdel~Khalek$^{\rm 114}$,
A.A.~Abdelalim$^{\rm 48}$,
O.~Abdinov$^{\rm 10}$,
R.~Aben$^{\rm 104}$,
B.~Abi$^{\rm 111}$,
M.~Abolins$^{\rm 87}$
% ...
\bigskip

% ...
$^{10}$ Institute of Physics, Azerbaijan Academy of Sciences, Baku, Azerbaijan\\
$^{11}$ Institut de F\'{i}sica d'Altes Energies and Departament de F\'{i}sica de la Universitat Aut\`{o}noma de Barcelona and ICREA, Barcelona, Spain\\
% ...
$^{20}$ Physikalisches Institut, University of Bonn, Bonn, Germany\\
% ...
$^{47}$ Fakult\"{a}t f\"{u}r Mathematik und Physik, Albert-Ludwigs-Universit\"{a}t, Freiburg, Germany\\
$^{48}$ Section de Physique, Universit\'{e} de Gen\`{e}ve, Geneva, Switzerland\\
% ...
$^{87}$ Department of Physics and Astronomy, Michigan State University, East Lansing MI, United States of America\\
% ...
$^{104}$ Nikhef National Institute for Subatomic Physics and University of Amsterdam, Amsterdam, Netherlands\\
% ...
$^{110}$ Homer L. Dodge Department of Physics and Astronomy, University of Oklahoma, Norman OK, United States of America\\
$^{111}$ Department of Physics, Oklahoma State University, Stillwater OK, United States of America\\
% ...
$^{114}$ LAL, Universit\'{e} Paris-Sud and CNRS/IN2P3, Orsay, France\\
% ...
\end{flushleft}

\end{document}